\documentclass[sigconf, nonacm]{acmart}

\usepackage[ruled,linesnumbered]{algorithm2e}
\newcommand\window{w}
\newcommand\graph{\mathcal{G}}

\SetKwComment{Comment}{$\triangleright$\ }{}

\begin{document}
\title{Low-Latency Sliding Window Connectivity}

\author{Chao Zhang}
\affiliation{%
  \institution{University of Waterloo}
  \city{Waterloo}
  \country{Canada}}
\email{chao.zhang@uwaterloo.ca}

\author{Angela Bonifati}
\affiliation{%
  \institution{Lyon 1 University, CNRS \& IUF}
  \city{Lyon}
  \country{France}}
\email{angela.bonifati@univ-lyon1.fr}

\author{M. Tamer Özsu}
\affiliation{%
  \institution{University of Waterloo}
  \city{Waterloo}
  \country{Canada}}
\email{tamer.ozsu@uwaterloo.ca}

\begin{abstract}
Connectivity queries, which check whether vertices belong to the same connected component, are fundamental in graph computations. 
Sliding window connectivity processes these queries over sliding windows, facilitating real-time streaming graph analytics. 
However, existing methods struggle with low-latency processing due to the significant overhead of continuously updating index structures as edges are inserted and deleted.
We introduce a novel approach that leverages  spanning trees to efficiently process queries. The novelty of this method lies in its ability to maintain spanning trees efficiently as window updates occur. Notably, our approach completely eliminates the need for replacement edge searches, a traditional bottleneck in managing spanning trees during edge deletions.
We also present several optimizations to maximize the potential of spanning-tree-based indexes.
Our comprehensive experimental evaluation shows that index update latency in spanning trees can be reduced by up to $458\times$ while maintaining query performance, leading to an $8\times$ improvement in throughput. Our approach also significantly outperforms the state-of-the-art in both query processing and index updates. Additionally, our methods use significantly less memory and demonstrate consistent efficiency across various settings.
\end{abstract}

\maketitle

\section{INTRODUCTION}\label{sec:intro}
Connectivity queries that check whether vertices belong to the same connected components (CCs) are fundamental to graph computations \cite{10.1145/3186728.3164139,sahu2020ubiquity}.
With the prevalence of graph representations in numerous real-world domains \cite{Newman2010,SakrBVIAAAABBDV21}, these queries are critical for a variety of applications, including social networks  \cite{10.1177/016555150202800601}, transport networks \cite{von2009public}, financial networks \cite{amy2023}, and more.

Modern data-driven applications, designed for efficient stream processing \cite{10.1145/2588555.2595641,10.1145/2723372.2742788,10.14778/2536222.2536229,10.1145/2517349.2522737,carbone2015apache}, necessitate the computation of connectivity queries over streaming graphs that are modeled as an unbounded and continuous flow of timestamped edges. 
This capability is essential for enabling real-time, low-latency data analysis.
In streaming graph processing, connectivity queries are typically computed in sliding windows \cite{10.1145/776985.776986} that contain the most recent edges and are updated as new edges arrive. 
A sliding window is defined by a window size $\alpha$ and a slide interval $\beta$. 
They are usually defined in time units, resulting in \textit{time-based sliding windows}.
These windows incorporate edges with timestamps that fall within the window's range. As streaming edges continuously arrive, the window content is updated by deleting \textit{expired edges} and inserting \textit{new edges}. The problem of computing connectivity queries within these sliding windows over a streaming graph is known as \textit{sliding window connectivity} \cite{zhang2024incremental}.

\begin{figure*}
    \centering
    \resizebox{0.9\textwidth}{!}{\includegraphics{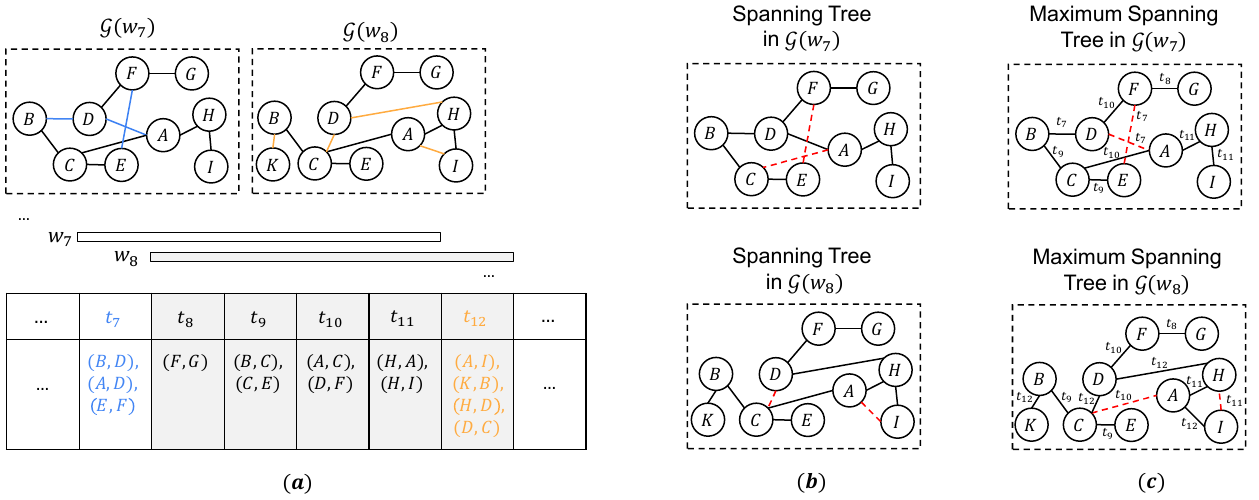}}
    \caption{Running example of sliding window connectivity.}
    \label{fig:swc}
\end{figure*}

\textit{Example}.
Figure \ref{fig:swc}(a) presents a running example of sliding windows with a window size of $5$ timestamps and a  slide interval of $1$ timestamp, where each timestamp could represent a time unit such as a minute. It shows window snapshot $\window_7$ containing edges with timestamps from $t_7$ to $t_{11}$ and illustrates the corresponding graph $\graph(\window_7)$.
When the window is moving from $\window_7$ to $\window_8$, expired edges (with timestamp $t_7$) will be deleted from $\graph(\window_7)$ and new edges (with timestamp $t_{12}$) are inserted, generating $\graph(\window_8)$.
The query result for vertices $I$ and $G$ in both $\graph(\window_7)$ and $\graph(\window_8)$ are True, as they belong to the same CC in $\graph(\window_7)$ and $\graph(\window_8)$, respectively.

A real-world example \cite{amy2023} involves tracking connectivity in a streaming graph of a financial network, where vertices represent users and edges represent transactions. A time-based sliding window focuses on recent activity, such as the past 30 minutes, continuously updating the graph by deleting expired edges (older transactions) and inserting new edges as transactions occur. This ensures that only the most recent interactions are analyzed, allowing detection of sudden bursts of connectivity between previously unconnected users. Such patterns can signal potential fraud, like money laundering, enabling real-time detection and response.


A graph-scan approach to sliding window connectivity, like applying Depth-First Search (DFS) for each query in every window snapshot, is highly inefficient due to the need for on-the-fly recomputation. This method fails to meet the high throughput and low latency demands of stream processing.

Alternatively, an index-based strategy can be used that involves using data structures designed for Fully Dynamic Connectivity (FDC) \cite{10.1145/945394.945398,10.1145/225058.225269,10.1145/276698.276715,10.1145/320211.320215,10.1145/265910.265914,10.1145/502090.502095,10.1145/335305.335345,10.1137/S0097539797327209,1.9781611973105.126,10.5555/2627817.2627898,kejlbergrasmussen_et_al:LIPIcs:2016:6395,10.5555/3039686.3039718,10.1145/3055399.3055415,10.1137/S0097539705447256,henzinger1998lower,10.14778/3551793.3551868,10.1145/335305.335345}.
These techniques maintain spanning trees within windows, with one tree per CC, supporting query processing, edge insertion, and edge deletion. Connectivity queries for $(u,v)$ can be efficiently answered by checking whether $u$ and $v$ belong to the same spanning tree, which involves simply traversing from each node to its root. New edge insertions can also be processed efficiently by linking spanning trees when the two endpoints of the new edge are not already connected.
However, significant performance challenges arise when updating spanning trees due to the deletion of expired edges.
Deleting an edge from a spanning tree requires searching for \textit{replacement edges} (see \S \ref{sec:replacement_edges}) to reconnect the resulting subtrees, which require a full graph traversal in the worst-case scenario.
In the running example, Figure \ref{fig:swc}(b) depicts the spanning tree in $\graph(\window_7)$, and the expired edge $(B,D)$ has a replacement edge $(E,F)$.

Another index-based strategy is Bidirectional Incremental Connectivity (BIC) \cite{zhang2024incremental}, built on Union-Find Trees (UFTs). Unlike the spanning trees used in FDC, UFTs support only edge insertion and query processing. To dynamically capture connectivity in windows with expired edge deletion and new edge insertion, BIC maintains both backward and forward UFTs, where the backward UFTs trace against the timeline and the forward UFTs follow it. These two UFTs are then merged to compute connectivity queries (see \S \ref{sec:related_works}). 
Although BIC avoids expired edge deletion, its major limitation lies in the backward computations of Union-Find Trees (UFTs), which run against the timeline, blocking query processing and the insertion of new streaming edges, resulting in high tail latency. More critically, as an index-based approach, BIC struggles to scale with large workloads due to the overhead involved in merging results from backward and forward UFTs to determine query results.

As noted, the main weakness of existing index-based approaches is their inefficiency in handling both query processing and index updates. FDC approaches use spanning trees for efficient queries but struggle with high index update latency due to replacement edge searches. BIC avoids these searches but requires costly backward UFT computations, impacting query efficiency.
The core challenge is designing an index that supports both efficient updates and queries, which is the focus of this work—a crucial need in stream processing, where minimizing latency is essential.

In this work, we propose a novel index-based framework that enables low-latency stream processing by efficiently handling both query processing and index updates. The framework can integrate FDC techniques while completely eliminating the need for replacement edge searches during edge deletions from spanning trees. As a result, queries are processed efficiently using spanning trees, and index updates are streamlined by removing the primary performance bottleneck of spanning trees, \textit{i.e.}, replacement edge searches.

The main idea of our framework is that a \textit{maximum spanning tree} is maintained for each CC in each window, using edge timestamps as weights, allowing for the straightforward deletion of expired edges. 
For instance, two spanning trees for tracking the CCs in $\graph(w_7)$ are shown in Figures \ref{fig:swc}(b) and (c).
When $w_7$ is transitioning to $w_8$, deleting the expired edge $(B,D)$ in the spanning tree in Figure \ref{fig:swc}(b) results in checking the existence of a non-tree edge (dotted in red) that can reconnect vertices $B$ and $D$, \textit{e.g.}, edge $(E, F)$, which is an expensive operation due to required graph traversal. 
The deletion in the maximum spanning tree in Figure \ref{fig:swc}(c), on the other hand, does not require such expensive computation or any additional operations besides the deletion itself.

Based on the framework, we propose novel indexing approaches for sliding window connectivity queries by integrating FDC techniques, ranging from basic spanning trees to more advanced methods such as D-Tree \cite{10.14778/3551793.3551868} and Link-Cut Tree \cite{10.1145/800076.802464}. FDC techniques are inherently complex due to the delicate balance they strike between edge insertion, edge deletion, and query processing. Integrating these techniques into the framework to create high-performance indexing methods is a challenging task, as the integration must simultaneously ensure efficient performance for all three operations.
Our key optimization technique is that, by eliminating the need for replacement edge searches, we can simplify the complex operations within FDC techniques, resulting in both low-latency processing and reduced memory consumption.
More importantly, this optimization is a general solution that can be applied to multiple FDC techniques rather than being limited to a specific one.
We then explore in detail the opportunities for optimizing the performance of each integrated FDC data structure within our framework.

Our main contributions are summarized as follows:
\begin{itemize}
    \item 
    We introduce a novel and efficient index-based approach for computing sliding window connectivity. This modular approach allows integration with various FDC data structures, leveraging their query processing strengths while addressing the performance bottleneck of index maintenance during edge expiration.

    \item 
    Our framework integrates various FDC data structures, identifying both universal and specialized optimizations to maximize performance and minimize memory usage.

    \item We present the first $O(\log{n})$ amortized-time approach for both query processing and index updates for sliding window connectivity.

    \item 
    Our methods reduce query latency by $1172\times$, window management latency by $13\times$, and improve throughput by $80\times$ compared to the most recent approach BIC. Compared to FDC approaches, we reduce window management latency by $458\times$ and improve throughput by $8\times$, while maintaining similar query latency.
    Additionally, our framework consumes significantly less memory and consistently delivers better results across various settings, including the number of queries processed, window size, and slide intervals.
\end{itemize}
%

\section{RELATED WORK}\label{sec:related_works}
\textit{Fully dynamic connectivity}.
FDC \cite{10.1145/945394.945398,10.1145/225058.225269,10.1145/276698.276715,10.1145/320211.320215,10.1145/265910.265914,10.1145/502090.502095,10.1145/335305.335345,10.1137/S0097539797327209,1.9781611973105.126,10.5555/2627817.2627898,kejlbergrasmussen_et_al:LIPIcs:2016:6395,10.5555/3039686.3039718,10.1145/3055399.3055415,10.1137/S0097539705447256,henzinger1998lower,10.14778/3551793.3551868,10.1145/335305.335345,10.1145/3555806} involves maintaining connectivity information over graphs undergoing both edge insertions and deletions. Data structures designed for computing FDC support three operations: \texttt{insert}, \texttt{delete}, and \texttt{query}. Spanning trees are maintained internally, with each processed edge classified as either a tree edge or a non-tree edge. Both insertion and deletion operations for either type of edges can be efficiently computed, except for the deletion of a tree edge, which requires a replacement edge search (see \S \ref{sec:replacement_edges}). This search can take $O(|V|+|E|)$ time in the worst case and is a major performance bottleneck of all existing FDC approaches.
While FDC data structures can be used to compute sliding window connectivity (as they inherently support \texttt{insert}, \texttt{delete}, and \texttt{query}), the performance bottleneck becomes more severe in this case because every edge inserted into the window will eventually expire and need to be deleted. In this work, we design a framework that leverages the benefits of FDC data structures for query processing while avoiding the expensive replacement edge search. 
Our experiments show that integrating D-Tree, the most efficient FDC solution in practice, with our framework significantly outperforms  vanilla D-Tree \cite{10.14778/3551793.3551868}.

\textit{Sliding window connectivity}. 
Vanilla FDC approaches, such as using D-Tree directly, can compute sliding window connectivity; however, their performance is bottlenecked by the need to search for replacement edges. In this work, we design frameworks  that alleviate these computationally expensive tasks in FDC approaches  (see the beginning of \S \ref{sec:related_works}).
The most recent approach for computing sliding window connectivity is BIC \cite{zhang2024incremental}, which supports \texttt{insert} and \texttt{query} operations while completely avoiding \texttt{delete} operations. 
It achieves this by computing two types of buffers for every disjoint chunk of  $\alpha$ timestamps: a forward buffer (along the timeline) and a backward buffer (against the timeline). Each type of buffer utilizes tailored Union-Find Trees \cite{10.1145/321879.321884,TARJAN1979110} to support \texttt{insert} and \texttt{query} operations. For connectivity information across forward and backward buffers, BIC constructs a bipartite graph and applies online traversal for query processing. In this sense, BIC serves as a partial index for sliding window connectivity.
Our experimental study identified two shortcomings of BIC. 
First, the computation of backward buffers, which involves scanning streaming edges over $ \alpha$ timestamps against the timeline, can result in high tail latency. 
Second, BIC's query processing, which involves traversing the bipartite graph, incurs very high latency with a large number of queries, limiting its scalability for large workloads.
In this work, we introduce a novel approach for computing sliding window connectivity based on spanning trees, in contrast to BIC, which relies on UFTs—entirely different data structures. As a result, our method eliminates the need for backward computation, a major source of overhead in BIC. While spanning trees have performance overhead, such as replacement edge searches during deletions, we develop novel techniques to fully eliminate this overhead while preserving their advantage of efficient query processing.

\textit{Minimum/maximum spanning trees in dynamic graphs}.
This problem is similar to FDC as both involve maintaining spanning trees in a graph with edge insertions and deletions. However, it has the added constraint that the total edge weight should be minimum (or maximum), which has been extensive studied \cite{10.1137/S0097539705447256,10.1145/502090.502095,10.1007/978-3-662-48350-3_62,10.5555/314161.314314,doi:10.1137/0214055,10.1007/978-3-540-72845-0_30,10.1145/3555806}.
The search for replacement edges is still required and becomes more challenging due to this weight constraint.
Our work focuses on sliding window connectivity and showcases a connection (see \S \ref{sec:mst}) among fully dynamic connectivity, maximum spanning trees, and sliding window connectivity by proposing a framework where timestamps of streaming edges are treated as edge weights. Importantly, we demonstrate that the replacement edge search can be entirely omitted when maintaining connectivity information in sliding windows over streaming graphs.
Theoretically, we introduce an $O(\log{n})$ amortized time per-operation approach based on the most asymptotically efficient dynamic tree, the Link-Cut Tree \cite{10.1145/800076.802464}. Practically, we extensively discuss optimization opportunities in implementing the advanced data structures. 
Our experimental findings align with previous studies \cite{CATTANEO2010404,10.1145/1498698.1594231} on FDC and minimum spanning trees, indicating that simple data structures, despite not being asymptotically efficient, can be fast in practice.


\textit{Stream processing systems}.
High throughput and low latency are crucial in stream processing \cite{10.1145/2588555.2595641,10.1145/2723372.2742788,10.14778/2536222.2536229,10.1145/2517349.2522737,carbone2015apache}. 
The continuous stream processing model, used in systems like Apache Flink \cite{carbone2015apache}, processes queries in real-time as windows complete. We consider this model for processing sliding window connectivity queries.
Although various streaming systems for graph computations exist \cite{10.1145/3302424.3303974,10.1145/3267809.3267811,10.1145/2960414.2960419,10.1007/978-3-319-43659-3_24,10.1145/3364180,6408680,10.1145/1526709.1526856,10.1007/978-3-642-25073-6_24,10.1007/978-3-642-17746-0_7,10.1007/978-3-319-25639-9_48,10.1007/s00778-021-00667-4}, they lack specific indexing methods for connectivity query. We aim to address this gap and efficiently support these fundamental graph computations.

\section{PROBLEM STATEMENT}
\subsection{Sliding Window Connectivity}
\textit{Connectivity queries}\label{sec:con-query}.
Consider an undirected graph  $\graph = (V, E)$ with a vertex set $V$ and an edge set $E$. 
A connectivity query $Q_c(u, v)$ in $\graph$ determines whether there is an undirected path between vertices $u$ and $v$. The result is True if and only if such a path exists.

\textit{Streaming graphs}.
A streaming graph is an infinite sequence of edges, $(e_1, e_2, \ldots)$, where each edge $e = (u, v, t)$ is undirected with endpoints $u$ and $v$ (referred to as $e.u$ and $e.v$) and a timestamp $t$ (referred to as $e.t$). For simplicity, we refer to an edge as $(u, v)$ when its timestamp is clear from context. We assume edges arrive in timestamp order, meaning for any edges $e_i$ and $e_j$, if $i < j$, then $e_i.t \leq e_j.t$ and $e_i$ arrives before $e_j$.

\textit{Streaming graphs in sliding windows}.
A sliding window maintains a fixed-size subset of a continuous data stream, denoted as $\window$. 
In this paper, we use a \textit{time-based window} model with a fixed \textit{window size} ($\alpha$) and \textit{slide intervals} ($\beta$), both in time units. 
This defines a sequence of window snapshots $(\window_1, \window_2, \ldots)$, where each $\window_i$ has a beginning timestamp $\window_i.t_b$ and an ending timestamp $\window_i.t_e$, containing $\alpha$ time units. For adjacent snapshots $\window_i$ and $\window_{i+1}$, $\window_{i+1}.t_b = \window_{i}.t_b + \beta$ and $\window_{i+1}.t_e = \window_{i}.t_e + \beta$. Each  $\window$ contains all edges $e_i$ where $\window.t_b \leq e_i.t \leq \window.t_e$, and the graph in $\window$ is denoted as $\graph(\window)$.
Since $\alpha$ is larger than $\beta$ in the context of sliding windows, window snapshots overlap.

\textit{Sliding window connectivity}.
Given a streaming graph $\graph$ and a time-based sliding window defined by a window size $\alpha$ and a slide interval $\beta$, the sliding window connectivity computes the connectivity query $Q_c(u, v)$  between vertices $u$ and $v$ in $(\graph(\window_1), \graph(\window_2), ...)$, where $(\window_1, \window_2, \ldots)$ is the sequence of the window snapshots defined by the sliding window.
In Figure \ref{fig:swc}(a), a path exists between $G$ and $I$ in $\graph(\window_7)$ and $\graph(\window_8)$, making $Q_c(G,I)=\text{True}$ in $\window_7$ and $\window_8$.

A naive approach like DFS  recomputes each query for each window snapshot, leading to high query latency and redundant computations.
We  aim at developing an index capable of efficiently processing connectivity queries within sliding windows. 
The principal challenge lies in efficiently maintaining the index. 
Specifically, as the window slides, \textit{expired edges} must be deleted and \textit{new edges} inserted, as illustrated in Figure \ref{fig:swc}(a).
The crux of the problem, therefore, is to efficiently incorporate these window updates into the maintained index, while simultaneously providing efficient query processing.

\subsection{Replacement Edges}\label{sec:replacement_edges}
FDC data structures process connectivity queries without recomputing connectivity for each window snapshot by maintaining spanning trees in each $\graph(\window_i)$.
However, their performance is hindered by the need to handle expired edges.
Each edge of a graph is classified either as a \textit{tree edge}, if it is included in the spanning trees, or as a \textit{non-tree edge}, if it is not. 
Deleting a non-tree edge is straightforward, as it does not affect connectivity.
In contrast, deleting a tree edge splits the spanning tree into two subtrees, each containing $u$ and $v$, respectively, and it is necessary to search among the non-tree edges related to these two subtrees to determine if there exists a non-tree edge that can reconnect the two split subtrees. Such a non-tree edge is referred to as a \textit{replacement edge}.

\begin{example}
Figure \ref{fig:swc}(b) shows the spanning tree in $\graph(\window_7)$, which includes two non-tree edges, $(A,C)$ and $(E,F)$. As $\graph(\window_7)$ transitions to $\graph(\window_8)$, the tree edge $(B,D)$ expires and must be deleted. Removing $(B,D)$ splits the spanning tree into two subtrees with vertex sets $\{B,C,E\}$ and $\{D,F,G,A,H,I\}$. A replacement edge, $(A,C)$, is then identified to reconnect the subtrees.
\end{example}

Replacement edge search is computationally expensive, requiring $O(n+m)$ time to traverse the graph with $n$ vertices $m$ edges to verify the existence of such non-tree edges.
In sliding window connectivity, this  becomes a significant performance bottleneck as every edge inserted into the window will be deleted thereafter.
In this work, we design an indexing framework that can avoid searching replacement edges when using FDC data structures. 

\begin{figure*}
    \centering
    \resizebox{\linewidth}{!}{
    \includegraphics{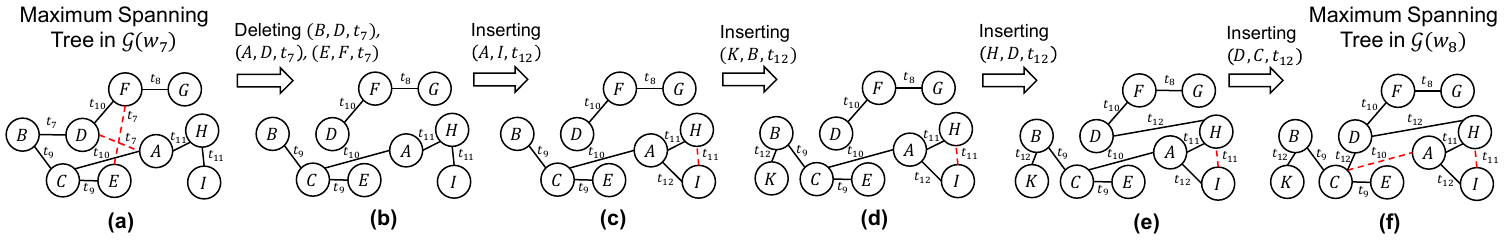}}
    \caption{Running example of maintaining a maximum spanning tree within window snapshots.}
    \label{fig:mst}
\end{figure*}

\section{FRAMEWORK}\label{sec:mst}
In this section, we introduce a framework that uses spanning trees to efficiently process sliding window connectivity queries, avoiding the need to search for replacement edges when deleting expired tree edges. Our approach maintains maximum spanning trees in sliding windows. 

\subsection{MSTs in Sliding Windows}
In our framework, each streaming edge $(u, v, t_i)$ is treated as a weighted edge $(u,v)$ labeled with the weight $t_i$ that is the timestamp of the streaming edge. 
We maintain one maximum spanning tree (MST) per connected component within each snapshot $\window_i$.

\begin{definition}[Maximum Spanning Tree]
Given a streaming graph $\graph$, and the sequence of window snapshots $(\graph(w_1),\graph(w_2), ... )$ of a time-based sliding window, for each connected component in each $\graph(w_i)$, an MST in $\graph(w_i)$ is a spanning tree such that the sum of edge weights of the spanning tree is the maximum among all possible spanning trees for the connected component.
\end{definition}

\begin{example}
Consider the running example in Figure \ref{fig:swc}. Since there is only one connected component in $\graph(\window_7)$, there will be a single MST. We illustrate two different spanning trees for this connected component with different total weights. The first spanning tree (Figure \ref{fig:swc}(b)) has $(A,C)$ and $(E,F)$ as non-tree edges, while the second has $(A,D)$ and $(E,F)$ as non-tree edges.  The total sum of edge weights in the first spanning tree is $73$ while the second has the sum of $76$. 
The second spanning tree with the higher total edge weight is the MST in $\graph(\window_7)$ depicted in Figure \ref{fig:swc}(c). 
\end{example}

\subsection{Deleting Expired Edges}
Once MSTs are maintained in sliding windows, each expired edge can be either a tree edge or a non-tree edge. Deleting a non-tree edge is straightforward. The following lemma demonstrates that with MSTs, deleting a tree edge is also straightforward.

\begin{lemma}\label{lemma:replacement_edge}
If an edge $e$ is deleted from an MST during the window movement from $w_i$ to $w_{i+1}$, then all replacement edges for $e$ will also be deleted during the same window movement from $w_i$ to $w_{i+1}$
\end{lemma}

In the MST framework, tree edges can be simply deleted without the need of searching for replacement edges,  taking only $O(1)$ time.
In contrast, FDC approaches requires $O(|V|+|E|)$ time for deleting a tree edge  because of searching for replacement edges.

\begin{example}
Consider the running example in Figure \ref{fig:swc}. When $\window_7$ transitions to $\window_8$, expired edges with timestamp $t_7$ need to be deleted. The first expired edge $(B,D)$ is a tree edge of the MST in $\graph(\window_7)$. Deleting $(B,D)$ does not require searching for replacement edges because $(A,D)$ and $(E,F)$, which are the candidates for replacement edges, also expire and are deleted along with $(B,D)$. Figure \ref{fig:mst}(b) depicts the MST after deleting these expired edges.
\end{example}

\subsection{Inserting New Edges}
Insertion of new streaming edges into a window requires more care since it is necessary to ensure that the MSTs constructed for the previous window remain MSTs after the insertions. This is especially true for non-tree edges.
Specifically, consider a new streaming edge $(u,v)$ with timestamp $t_i$. If $u$ and $v$ are not connected before the insertion (\textit{i.e.}, in the previous window), then edge $(u,v)$ is treated as a tree edge by linking the two subtrees containing $u$ and $v$. This can be accomplished by simply making $v$ the root of its tree and linking $v$ as a child of $u$, or vice versa. After this, the spanning tree containing $(u,v)$ remains an MST.
In the case where $u$ and $v$ are already connected, edge $(u,v)$ is treated as a non-tree edge. Since the timestamp $t_i$ of the new streaming edge $(u,v)$ is equal to or greater than the timestamp of any edge in the current window before the insertion, simply storing $(u,v)$ as a non-tree edge may violate the MST constraint. 
In this scenario, we apply additional computations based on the following lemma.

\begin{lemma}\label{lemma:min_edge}
When inserting a streaming new edge $(u,v)$ into an MST in any window snapshot of a sliding window, and $u$ and $v$ are already connected in the MST, the spanning tree will remain an MST by converting the minimum tree edge\footnote{If the minimum edges are not unique, any can be selected for  converting.} in the cycle containing $(u,v)$ into a non-tree edge and then inserting $(u,v)$ as a tree edge.
\end{lemma}

\begin{example}
Consider the example in Figure \ref{fig:mst}. When inserting $(A,I,t_{12})$ into the MST in Figure \ref{fig:mst}(b), $A$ and $I$ are already connected, creating a cycle containing $\{A,H,I\}$. In this cycle, both the weights of $(A,H,t_{11})$ and $(H,I,t_{11})$ are smaller than the weight of $(A,I,t_{12})$. Therefore, either of these edges can be changed into a non-tree edge, followed by changing $(A,I,t_{12})$ into a tree edge. Consequently, the spanning tree in Figure \ref{fig:mst}(c) is an MST.
For the subsequent edges $(K,B,t_{12})$ and $(H,D,t_{12})$, as their endpoints are not connected in Figure \ref{fig:mst}(c) and Figure \ref{fig:mst}(d), respectively, they are simply inserted as tree edges. Inserting the last new edge $(D,C,t_{12})$ will create a cycle. Therefore, the smallest edge in the cycle, $(A,C,t_{10})$, is changed into a non-tree edge, and $(D,C,t_{12})$ is eventually inserted as a tree edge, resulting in the MST shown in Figure \ref{fig:mst}(f).
\end{example}

Compared to vanilla spanning trees, the MST framework adds computation for converting minimum tree edges into non-tree edges, which can be leveraged to avoid searching replacement edges when deleting expired edges. This trade-off is highly beneficial, as finding minimum edges only requires traversing paths within the spanning trees, whereas deleting expired edges in vanilla trees require traversing the entire graph in the window to find replacement edges.

    \begin{algorithm}[]
    \SetAlgoVlined
    \SetKwFunction{KwInsert}{insert}
    \SetKwFunction{KwDelete}{delete}
    \SetKwFunction{KwQuery}{query}
    \SetKwProg{Pr}{procedure}{}{}
    \SetKwProg{Fn}{function}{}{}
        \Pr{\KwInsert{$e$}}{
            \uIf{$mst.query(e.u,e.v)$ is False}{
                $mst.insertTreeEdge(e)$\;
            }\Else{
                $e_{min} \leftarrow mst.findMinInCycle(e)$\;
            
                \uIf{$e_{min}.t < e.t$}{
                    $mst.deleteTreeEdge(e_{min})$\;
                    $mst.insertNonTreeEdge(e_{min})$\;
                    $mst.insertTreeEdge(e)$\;        
                }\Else{
                    $mst.insertNonTreeEdge(e)$\;
                }
            }
        }
        \Pr{\KwDelete{$e$}}{
            \uIf{$mst.isTreeEdge(e)$ is True}{
                     $mst.deleteTreeEdge(e)$ \Comment*[r]{\footnotesize{\textbf{no replacement edge search}}}
            }\Else{
                $mst.deleteNonTreeEdge(e)$\;
            }
        }
        \Fn{\KwQuery{$u,v$}}{
            \Return $mst.query(u,v)$\;
        }
    \caption{The MST Framework}\label{algo:mst}
    \end{algorithm}

\subsection{Sliding Window Connectivity with MSTs}
Query $Q_c(u,v)$ can be computed by checking whether $u$ and $v$ belong to the same MST. 
Additionally, for each new edge $(u,v)$, we determine if the new edge should be treated as a tree edge by verifying if $Q_c(u,v) = \text{False}$ before inserting the new edge. If not, then the new edge is a non-tree edge.

We present the computations of sliding window connectivity with MSTs in Algorithm \ref{algo:mst}. Specifically, line 14 addresses the case of deleting a tree edge from the MST, and according to Lemma \ref{lemma:replacement_edge}, there is no need to search for replacement edges. Lines 5-11 demonstrate the procedure for inserting a non-tree edge into the MST, implemented according to Lemma \ref{lemma:min_edge}.

Based on Lemmas \ref{lemma:replacement_edge} and \ref{lemma:min_edge}, we demonstrate in Theorem \ref{theorem:mst_swc} that Algorithm \ref{algo:mst} is correct, even though it omits the search for replacement edges when deleting tree edges.

\begin{theorem}\label{theorem:mst_swc}
Algorithm \ref{algo:mst} for computing sliding window connectivity with MSTs is sound and complete.
\end{theorem}

In the remainder of the paper, we describe how to use FDC data structures to implement the abstract data types of MSTs required in Algorithm \ref{algo:mst} in Section \ref{sec:implementation}. Then, we discuss optimized implementations in Section \ref{sec:optimized_implementaion} using the most advanced techniques.

\section{INTEGRATING FDC TECHNIQUES}\label{sec:implementation}
In this section, we explain how to integrate FDC techniques into our framework, enabling the application of Algorithm \ref{algo:mst} for computing sliding window connectivity. 
We then propose MST D-Tree, which is based on D-Tree \cite{10.14778/3551793.3551868}, the most efficient FDC solution in practice.

\subsection{Computing MSTs}\label{sec:computing_mst}
FDC data structures are designed to compute connectivity queries in graphs that undergo both edge insertions and deletions. These data structures support the following operations: 
\begin{itemize}
    \item \texttt{reRoot(v)}: Make a vertex $v$ the root of its tree.
    
    \item \texttt{link(u,v)}: Link a vertex  $u$ , which is the root of its spanning tree, as a child of another vertex $ v $. If $ u $ is not  the root of its tree, apply \texttt{reRoot(u)} before performing  \texttt{link(u,v)}.

\item \texttt{cut(v)}: Separate a vertex $ v $ from its parent.

\item \texttt{findRoot(v)}: Find the root of a vertex $ v $ in its spanning tree, enabling the computation of query $ Q_c(u,v) $ by checking whether \texttt{findRoot(u)} and \texttt{findRoot(v)} are the same.
\end{itemize}
With the above operations, all necessary steps of MST in Algorithm \ref{algo:mst} can be executed, except for the \texttt{findMinInCycle} operation.
Supporting this operation requires storing the edge weights. Since each child has only one parent in a spanning tree, the weight can be stored as an attribute of the child vertex for a given tree edge.
To support \texttt{findMinInCycle}, we can first compute the Lowest Common Ancestor (LCA) of $u$ and $v$ and then compute the minimum edge in the paths from $u$ and $v$ to their LCA. 


To summarize, any FDC data structure  can be adapted to compute  MST in Algorithm \ref{algo:mst}. 
It is important to note that FDC data structures can also be used to compute sliding window connectivity without adaption, as they inherently support  \texttt{insert}, \texttt{delete}, and \texttt{query} required for sliding window connectivity. We refer to such approaches as \textit{vanilla solutions}.
In contrast, integrating FDC data structures to compute MSTs in Algorithm \ref{algo:mst} for computing sliding window connectivity is referred to as \textit{MST solutions}. The main difference between the vanilla solution and the MST solution is that, in the former case, it is necessary to search for replacement edges when deleting tree edges, a step that the MST solution eliminates. Although the MST solution requires the additional \texttt{findMinInCycle} operation, it is significantly more efficient than the vanilla solution in terms of both throughput and latency, as demonstrated by our experimental evaluation. 

\subsection{D-Tree Integration}\label{sec:mst-dtree}
\textbf{Vanilla D-Tree} refers to the approach of using D-Tree directly to compute sliding window connectivity.
In this setting, each vertex $v$ stores the following information: 
the parent of $v$; 
a set of vertices that are the children of $v$; 
a set of vertices that have non-tree edges related to $v$, referred to as non-tree vertices; 
the number of vertices in the subtree rooted at $v$, defined as the \textit{size} of $v$. 
To optimize performance and reduce the average computational cost, D-Tree incorporates the following three techniques:\\
\textit{Technique 1.} Consider the path $(v,...,x,r)$ from vertex $v$ to its parent $r$, where $x$ is a child of $r$. If the size of $x$ exceeds half the size of $r$, then $x$ will be re-rooted to become the root of the tree. Specifically, $r$ will be designated as a child of $x$ following this re-rooting process.\\
\textit{Technique 2.} Consider a non-tree edge $(u,v)$, with $d_u$ and $d_v$ representing the distances from $u$ and $v$ to the root $r$, respectively. If the absolute difference in their distances, $|d_u-d_v|$, is not less than 2, then the tree structure is modified. Specifically, the ancestor of the vertex ($u$ or $v$) that is far from $r$, located $|d_u-d_v|-1$ hops away, is reclassified as a non-tree vertex. Subsequently, this far vertex is linked as a child of the vertex that is close to $r$.\\
\textit{Technique 3.} Consider the scenario where $(u,v)$ is a tree edge being deleted, with $u$ having a smaller subtree size than $v$. Let $(u',v')$ represent the replacement edge for $(u,v)$, where $u'$ and $v'$ are located in the subtrees containing $u$ and $v$, respectively. The selection of $(u',v')$ is made from all potential replacement edges such that the distance from $v'$ to its root is minimized.

\textbf{MST D-Tree} refers to the use of D-Tree to compute MSTs in Algorithm \ref{algo:mst}. Unlike the vanilla D-Tree, the MST D-Tree incorporates Technique 1 and 2, but not Technique 3. This is because MST D-Tree does not require searching for replacement edges, thus eliminating the need for the procedures outlined in Technique 3.
Additionally, to facilitate the \texttt{findMinInCycle} operation, each vertex $v$ in the MST D-Tree is enhanced to store a weight. This weight represents the weight of the edge from $v$ to its parent.

\section{OPTIMIZATION TECHNIQUES}\label{sec:optimized_implementaion}
In this section, we present an optimization for improving Algorithm \ref{algo:mst}. We integrate simple spanning trees, D-Tree, and Link-Cut Tree (LC-Tree) into an optimized framework, representing the most efficient data structures both practically and theoretically. We then propose three data structures based on this integration and discuss their optimization opportunities.

\subsection{Optimized MST}
We observe that in the MST framework, the information stored in each vertex can be simplified because MSTs do not require searching for replacement edges when deleting tree edges, as per Lemma \ref{lemma:replacement_edge}.
Specifically, a vertex $v$ does not need to store the set of its children and the set of non-tree edges associated with $v$. Consequently, the only necessary information for each vertex $v$ includes: the parent of $v$, the weight of the edge from $v$ to its parent, and the size of the subtree rooted at $v$.
This is sufficient to maintain minimal topological information about MSTs, enabling traversal from any vertex  $v$ to its root for processing queries. 
This streamlined approach proves to be highly effective in practice, as evidenced by our experimental evaluation. 
Throughout the remainder of the paper, we refer to the MST framework that incorporates this optimization as the Optimized MST (OMST). 
It is important to note that OMST can be integrated with any data structure designed for FDC.
The algorithm for computing sliding window connectivity with OMSTs largely mirrors Algorithm \ref{algo:mst}, with the exception that lines 8, 10, 11, 15, and 16 are omitted. These lines, which pertain to the handling of non-tree edges, are not necessary in the OMST framework.

\begin{figure*}
    \begin{minipage}{0.32\textwidth}
        \centering
        \resizebox{\linewidth}{!}{
        \includegraphics{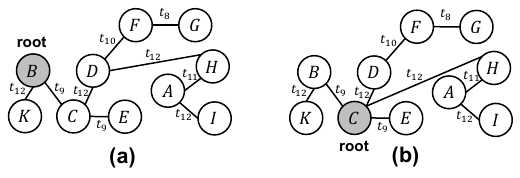}}
        \caption{OMST S-Tree and OMST D-Tree.}
        \label{fig:omst-dtree}
    \end{minipage}
    \hfill
    \begin{minipage}{0.65\textwidth}
        \centering
        \resizebox{\textwidth}{!}{
        \includegraphics{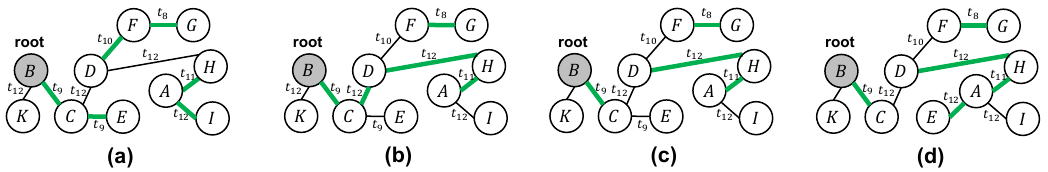}}
        \caption{Example of OMST LC-Tree, with preferred edges in green.}
        \label{fig:omst-lct}
    \end{minipage}  
\end{figure*}


\subsection{OMST S-Tree}\label{sec:omst-s-tree}
We first introduce an approach by integrating simple spanning trees into the OMST framework.
Consider inserting a tree edge $(u,v)$.
If the size of the root of $u$ is smaller than that of $v$, then re-root the tree containing $u$ and link $u$ as a child of $v$.
Otherwise, re-root $v$ and link $v$ as a child of $u$.
Deleting a tree edge involves cutting the child from its parent. A connectivity query $Q_c(u,v)$ can be processed by checking whether $u$ and $v$ share the same root.
For inserting a non-tree edge $(u,v)$, as depicted in Section \ref{sec:computing_mst}, the process includes finding the minimum edge in the cycle containing $(u,v)$, deleting this minimum tree edge, and inserting $(u,v)$ as a tree edge.
This streamlined method is referred to as OMST Simple Tree, or OMST S-Tree for short. 
As demonstrated in our experiments, employing this simple approach within the OMST framework proves to be highly efficient for computing sliding window connectivity.

\begin{example}\label{example:omst_stree}
    Consider the MST of $\graph(\window_8)$ in Figure \ref{fig:mst}(f). The corresponding OMST S-Tree is shown in Figure \ref{fig:omst-dtree}(a), where $B$ is the root in the tree. 
    Notice that the non-tree edges in  Figure \ref{fig:mst}(f) do not need to be stored in the OMST S-Tree shown in Figure \ref{fig:omst-dtree}(a). 
    Consider inserting edge $(C,H,t_{12})$ into the OMST S-Tree. As $Q_c(C,H) = \text{True}$, the edge is treated as a non-tree edge. In addition, the edge weight is not larger than the minimum edge in the cycle of $(C,H)$. Therefore, the edge does not need to be stored. Eventually, the OMST S-Tree remains unchanged after the insertion. 
\end{example}

\subsection{OMST D-Tree}\label{sec:omst-d-tree}
OMST D-Tree involves integrating D-Tree within the OMST framework. 
Similar to MST D-Tree, OMST D-Tree incorporates Techniques 1 and 2 from  Vanilla D-Tree but omits Technique 3, as OMST D-Tree does not require searching for replacement edges when deleting tree edges.
Compared to MST D-Tree, OMST D-Tree reduces unnecessary data storage by excluding non-tree edges. 
As prescribed by Technique 2, spanning trees may be restructured upon inserting a non-tree edge, contingent on specific conditions on vertex distances.
In OMST D-Tree, during the application of Technique 2, only tree edges are retained, and the corresponding insertion of non-tree edges is omitted.
In addition, Technique 2 is not applied when processing the minimum edge in the cycle containing $(u,v)$ during the insertion of $(u,v)$. 
This is necessary to ensure that the spanning trees remain MSTs, as Technique 2 might violate this condition.
We note that OMST D-Tree uses the approach depicted in Section \ref{sec:computing_mst} to identify the minimum edge, which is the same as Vanilla D-Tree and MST D-Tree.

\begin{example}\label{example:omst_dtree}
    Consider the MST of $\graph(\window_8)$ in Figure \ref{fig:mst}(f). The corresponding OMST D-Tree is shown in Figure \ref{fig:omst-dtree}(a), where $B$ is the root in the tree. 
    Consider inserting $(C,H,t_{12})$ into the OMST D-Tree. 
    To process the insertion, the first step is to evaluate $Q_c(C,H)$.
    With Technique 1 of D-Tree, OMST D-Tree identifies the subtree rooted at $C$ has more than half vertices than the tree. Therefore, OMST D-Tree applies the re-root operation to make $C$ the root of the tree.
    Then, since $Q_c(C,H) = \text{True}$,  $(C,H,t_{12})$ is treated as a non-tree edge. 
    At this step, since the weight of $(C,H)$ is not larger than the minimum edge weight in the cycle of $(C,H)$, Technique 2 of D-Tree is triggered because the distances of $C$ and $H$ are $0$ and $2$, respectively. Therefore,  $(D,H)$ is removed and $(C,H)$ is inserted. 
    Eventually, Figure \ref{fig:omst-dtree}(b) shows the OMST D-Tree after processing the insertion of $(C,H,t_{12})$.     
\end{example}

\subsection{OMST Link-Cut Tree}\label{sec:omst-lc-tree}
LC-Tree \cite{10.1145/800076.802464} is one of the most advanced data structures for maintaining FDC. 
The most interesting aspect of LC-Tree is that it can find the root $r$ of a vertex $v$ without traversing every edge in the path from $v$ to $r$. This is not possible with S-Tree or D-Tree. 
In general, vertices in the spanning tree are partitioned into paths, where each vertex belongs to only one path. Such paths are known as \textit{preferred paths}. 
To define the mechanism of forming a preferred path, LC-Tree provides an \texttt{access} operation. Specifically, whenever applying the \texttt{access} operation on a vertex $v$, the path from $v$ to its root $r$ will become a preferred path, where the child of $v$ is not included in the preferred path.
Each preferred path is stored using an auxiliary data structure that is usually a splay tree \cite{10.1145/800076.802464,10.1145/1498698.1594231} (keyed by the distances from vertices to their roots). 
With the \texttt{access} operation, the \texttt{findRoot} operation on $v$ can be executed by first accessing $v$, followed by visiting the leftmost vertex in the splay tree representing the path from $v$ to its root. 
The \texttt{link} and \texttt{cut} operation can also be efficiently executed based on the \texttt{access} operation \cite{10.1145/800076.802464}.

We denote the approach of integrating LC-Tree into the OMST framework as OMST Link-Cut Tree, or OMST LC-Tree for short. 
Similar to OMST S-Tree and OMST D-Tree, each vertex $v$ in LC-Tree is augmented to store the weight of the tree edge between $v$ and its parent. Then, we can compute the minimum edge of a path by augmenting vertices in  splay trees, representing paths in the spanning tree, to store aggregation information. 

We discuss two optimization opportunities in using OMST LC-Tree to compute sliding window connectivity below.

\textit{Optimization 1} focuses on processing the query $Q_c(u,v)$. The straightforward approach involves checking if \texttt{findRoot(u)} equals \texttt{findRoot(v)}. As previously discussed, the \texttt{findRoot} operation requires first forming the preferred path from a vertex to its root, followed by traversing the splay tree representing this preferred path. However, we can avoid the splay tree traversal when processing queries.
Specifically, for the query $Q_c(u,v)$, we first apply \texttt{access(u)} and then \texttt{access(v)}. Following these operations, $Q_c(u,v)=\text{False}$ if the preferred path from $u$ to its root remains unchanged. This can be verified in constant time by checking the pointers associated with $u$. This indicates that the root of $v$ differs from the root of $u$. Conversely, if the preferred path is altered, $Q_c(u,v)=\text{True}$. 
We note that this LC-Tree optimization technique can also be applied out of the OMST framework.

\textit{Optimization 2} pertains to computing the minimum edge in the cycle containing an edge $(u,v)$. 
The solution, described in Section \ref{sec:computing_mst}, involves traversing the entire paths from 
$u$ and $v$ to their root to determine their LCA, followed by traversing from $u$ and $v$ to the LCA.
We observe that this process can be simplified by carefully applying the \texttt{access} operations.
To achieve this, we augment the \texttt{access} operation as follows: \texttt{access(v)} will return the parent of the last edge (counting from $v$ to $v$'s root) that was not previously included in the preferred path from $v$ to its root. In this way, the LCA of $u$ and $v$ can be computed by first applying \texttt{access(u)} and then using the returned value from \texttt{access(v)} as their LCA.
After determining the LCA, to compute the minimum edge in the path from $u$ to the LCA, we first apply \texttt{access(u)} and then \texttt{access(LCA)}. 
This ensures that the minimum edge of the path from $u$ to the LCA can be found as aggregation information in a single splay tree.
The minimum edge from $v$ to the LCA can be computed in the same manner.

\begin{example}\label{example:omst_lctree}
Consider the MST in $\graph(\window_8)$ shown in Figure \ref{fig:mst}. We explain how OMST LC-Tree operates when inserting the edge $(E,A,t_{12})$ into the MST.
First, we determine whether $(E,A)$ should be treated as a tree edge by executing $Q_c(E,A)$. As discussed in Optimization 1, we first apply \texttt{access(E)}, followed by \texttt{access(A)}. Figures \ref{fig:omst-lct}(a) and \ref{fig:omst-lct}(b) depict the OMST LC-Tree after \texttt{access(E)} and \texttt{access(A)}, respectively. Since the preferred path from $E$ to the root $B$ has  changed, we conclude $Q_c(E,A)=\text{True}$. This indicates that $(E,A)$ should be treated as a non-tree edge.
Next, we compute the minimum edge in the cycle containing $(E,A)$. Following Optimization 2, we first take the returned value of \texttt{access(A)}, which is $C$. This is because two edges, $(H,D)$ and $(D,C)$, were not previously included in the preferred path from $A$ to $B$, with $(D,C)$ being the last one encountered when traversing from $A$ to $B$.
Then, \texttt{access(C)} will make $(D,H,A)$ a single splay tree, as shown in Figure \ref{fig:omst-lct}(c). This splay tree can provide the minimum edge from $A$ to $C$ (noting that edge weights are stored in child vertices, so the weight of $(D,C)$ is stored in $D$), which is $(A,H,t_{11})$. The minimum edge from $E$ to the LCA $C$ can be computed similarly, which is $(E, C, t_{9})$.
Finally, since $(E, C, t_{9})$ is smaller than $(E,A,t_{12})$, we cut $E$ from $C$ and link it as a child of $A$. Figure \ref{fig:omst-lct} shows the OMST LC-Tree after processing the insertion of $(E,A,t_{12})$.
\end{example}

\subsection{Complexity Analysis}
Let $n$ and $m$ be the number of vertices and edges in a window. 

OMST S-Tree has \(O(n)\) worst-case and amortized time complexity for \texttt{query}, \texttt{insert}, and \texttt{delete} due to traversing paths from vertices to their roots. Deletion is also \(O(n)\) because it involves updating vertex sizes along these paths.
OMST D-Tree and MST D-Tree have the same worst-case time complexity as OMST S-Tree for the same reason. 
However, OMST D-Tree and MST D-Tree are expected to have better average-case complexity due to applying D-Tree's Techniques 1 and 2, reducing the path length from vertices to their roots on average (see D-Tree \cite{10.14778/3551793.3551868} for details).
OMST LC-Tree can achieve $O(\log{n})$ amortized time for each of \texttt{query}, \texttt{insert}, and \texttt{delete} because all these operations are executed by a constant number of invocations of the \texttt{access} operation, taking $O(\log n)$ amortized time \cite{10.1145/800076.802464,10.1145/1498698.1594231} when using splay trees to store preferred paths (see LC-Tree \cite{10.1145/800076.802464} for details).

MST D-Tree takes $O(n+m)$ space as both vertices and all edges are stored. 
In contrast, each of OMST S-Tree, OMST D-Tree, and OMST LC-Tree requires only $O(n)$ space as only vertices and tree edges are stored.

\section{EXPERIMENTAL EVALUATION}
We evaluate the proposed approaches against a number of existing methods.
Our experiments are dedicated to analyzing the proposed approaches by using various metrics such as throughput, query latency, window management latency (including index maintenance), the impact of workload size, the impact of window size and slide interval, and memory usage.


\textit{Result summary}. First, compared to Vanilla D-Tree \cite{10.14778/3551793.3551868}, our approaches achieve similar query latency but demonstrate up to a $458\times$ reduction in window management latency, highlighting the effectiveness of avoiding replacement edge searches when deleting expired edges in our framework, along with an $8\times$ improvement in throughput. 
Second, compared to the most recent baseline, BIC \cite{zhang2024incremental}, our methods show up to a $1172\times$ reduction in query latency, a $13\times$ reduction in window management latency, and an up to $80\times$ improvement in throughput. Importantly, our methods consistently outperform existing approaches across various configurations, including different window sizes, slide intervals, and workload sizes, while using significantly less memory.

\subsection{Experimental Setup}\label{sec:exp_setup}
The following $8$ approaches are compared in our experimental evaluation: \textbf{Depth-First Search (DFS)}, \textbf{RWC}, \textbf{Vanilla D-Tree} \cite{10.14778/3551793.3551868}, \textbf{BIC} \cite{zhang2024incremental}, \textbf{MST D-Tree}, \textbf{OMST S-Tree}, \textbf{OMST D-Tree},  and \textbf{OMST LC-Tree}. Among these, DFS and RWC are non-indexing approaches, while the others are indexing approaches.
The DFS approach applies depth-first search to compute each query in each window. 
RWC recomputes the connected components in each window and uses them to process queries.
Vanilla D-Tree uses D-Tree as the FDC data structure to compute sliding window connectivity. 
We note that other FDC data structures, \textit{e.g.}, Euler-Tour Tree \cite{10.1137/0214061} and HDT \cite{10.1145/276698.276715,10.1145/502090.502095}, are excluded from our experimental evaluation due to their suboptimal performance compared to BIC and Vanilla D-Tree, as demonstrated previously \cite{zhang2024incremental}.
MST D-Tree (\S \ref{sec:mst-dtree}), OMST S-Tree (\S \ref{sec:omst-s-tree}), OMST D-Tree (\S \ref{sec:omst-d-tree}), and OMST LC-Tree (\S \ref{sec:omst-lc-tree}) are the approaches proposed in this paper. 
Our codebase\footnote{Open Source Link: \url{https://github.com/chaozhang-cs/lswc}} for implementation and experimental evaluation has been made publicly available online.

\begin{table}[h]
    \centering
    \caption{Overview of datasets.}
    \label{table:datasets}
    \resizebox{0.46\linewidth}{!}{
    \begin{tabular}{|c|r|r|r|r|}
    \hline
         \textbf{Dataset}   & $|V|$ & $|E|$ &AD &D \\ \hline
         \href{http://konect.cc/networks/youtube-u-growth/}{Youtube-growth (YG)}& 3.2M & 14.4M & $5.2$ & $31$\\ \hline
          \href{https://snap.stanford.edu/data/wiki-topcats.html}{Wiki-top (WT)}     & 1.7M  & 28.5M & - & 9 \\ \hline           
          \href{http://konect.cc/networks/soc-pokec-relationships/}{Pokec (PR)}        & 1.6M  & 30.6M &4.6& 14\\ \hline
          \href{https://snap.stanford.edu/data/com-LiveJournal.html}{LiveJournal (LJ)}  & 3.9M  & 34.6M &-& 17\\ \hline
          \href{http://konect.cc/networks/sx-stackoverflow/}{StackOverflow (SO)}& 2.6M  & 63.4M &3.9 & 11 \\ \hline
        \end{tabular}}
        \resizebox{0.5\linewidth}{!}{
        
        \begin{tabular}{|c|r|r|r|r|}
        \hline
             \textbf{Dataset}   & $|V|$ & $|E|$ &AD&D\\ \hline        
         \href{http://konect.cc/networks/orkut-links/}{Orkut (OR)}           & 3M    & 117.1M&4.2&10 \\ \hline          
         \href{https://github.com/ldbc/ldbc_snb_bi/blob/main/snb-bi-pre-generated-data-sets.md}{LDBC SNB Knows (LK)}   & 3.3M  & 187.2M & - & -\\ \hline
          \href{https://ldbcouncil.org/benchmarks/graphalytics/}{Graph-500  (GF)}       & 17M   & 523.6M &-&-\\ \hline          
          \href{https://snap.stanford.edu/data/com-Friendster.html}{Friendster (FS)}   & 63.6M & 1.8B &-& 32  \\ \hline
          \href{https://www.semanticscholar.org/product/api}{Semantic Scholar (SC)} & 65M & 8.27B & - &- \\ \hline
    \end{tabular}}
\end{table}

\textit{Datasets and workloads}.
We use $10$ datasets listed in Table \ref{table:datasets}, comprising real-world graphs \cite{snapnets}—namely, YG, WT, PR, LJ, SO, OR, FS, and SC—and synthetic graphs from industrial-grade benchmarks: LK \cite{10.14778/3574245.3574270} with the scale factor $1000$, and GF \cite{murphy2010introducing} with the scale factor $25$. 
The average distance (AD) and diameter (D) are also shown in Table \ref{table:datasets} for those such that their AD and D are available online. 

\textit{Streaming graphs}.
Each dataset is used to simulate a streaming graph by  assigning a timestamp to each edge using a uniform distribution. This is not done for LK and SO, since they already have timestamps.

\textit{Evaluation metrics}.
Our experiments report throughput and tail latency, key metrics in stream processing systems.
Throughput is calculated by dividing the processing time of a dataset by the number of edges within that dataset. 
For tail latency, we measure query processing time and window management time at the completion of each window snapshot, referred to as query latency and window management latency. Once a window is complete, queries are computed first, followed by updating the window by removing expired edges.
In non-indexed approaches such as DFS and RWC, query latency refers to the time taken to compute queries using the graph in the current window, while window management latency reflects the time spent deleting expired edges from the current window.
In contrast, for index-based approaches such as BIC, Vanilla D-Tree, MST D-Tree, OMST S-Tree, OMST D-Tree,  and OMST LC-Tree, query latency refers to the time required to compute queries utilizing a specific index.
Window management latency of index-based approaches except BIC encompasses the removal of expired edges both from the current window and the index. 
BIC does not perform deletion from the window and the index; instead, it involves backward computation over the entire chunk when the chunk, which matches the window size, is full—this happens once every $\alpha / \beta$ windows. 
The backward computation is the only overhead for window management in BIC, and it is considered as its window management latency \cite{zhang2024incremental}.
For both query and window management, we report the 95th percentile (P95) and 99th percentile (P99) latency results to offer a thorough understanding of index performance.
While P95 captures performance for the majority of cases, P99 provides critical insight into the system's behavior under worst-case scenarios, ensuring potential bottlenecks are identified and addressed.

\textit{Settings}. 
Queries and window updates (including edge insertions and deletions) are performed in sliding windows, with the number of edges and slide intervals impacting performance. We explore various settings for these parameters.
Specifically, we begin the experiments in \S \ref{sec:performance} with a window size and a slide interval such that, on average, each window contains $3$M edges and each slide interval contains $150$K edges.
In \S \ref{sec:workload}, we examine the impact of the number of queries in the workloads, considering $1$, $10^2$, $10^4$, and $10^6$ queries, with each window containing $20$M edges and each slide interval containing $1$M edges.
In \S \ref{sec:scalability}, we assess the impact of window sizes and slide intervals under the following two scenarios: a fixed slide interval with each slide containing on average 1M edges, and varying window sizes with each window containing on average  10M, 20M, 40M, and 80M edges, respectively; a fixed window size with each window containing on average 80M edges, and varying slide intervals with each slide containing on average 1M, 2M, 4M, and 8M edges, respectively.
We use all datasets in \S \ref{sec:performance}, and then focus on GF, FS, and SC in \S \ref{sec:workload} and \S \ref{sec:scalability} as they have a significant number of edges, enabling us to assess larger window sizes and slide intervals.
In \S \ref{sec:performance} and \S \ref{sec:scalability}, we compute a workload of $10^5$ queries, comparing all approaches except DFS due to its suboptimal performance. DFS is included in \S \ref{sec:workload} with smaller workloads.
We report throughput, query latency, and window management latency in all these experiments.
At last, we evaluate the memory consumption of all approaches in \S \ref{sec:memory_udage} using these settings of window sizes and slide intervals.

All the experiments are run on a server with Ubuntu 22.04, $160$ CPUs of $2.30$GHz, and $1$TB main memory.

\begin{figure*}
\centering
\resizebox{0.25\textwidth}{!}{
\includegraphics{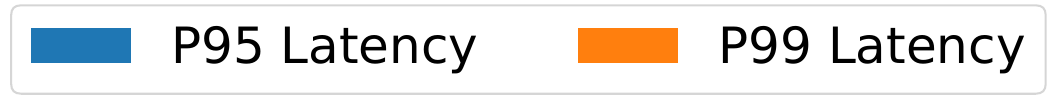}
} \\
\begin{minipage}{0.094\linewidth}
\resizebox{\textwidth}{!}{
    \includegraphics{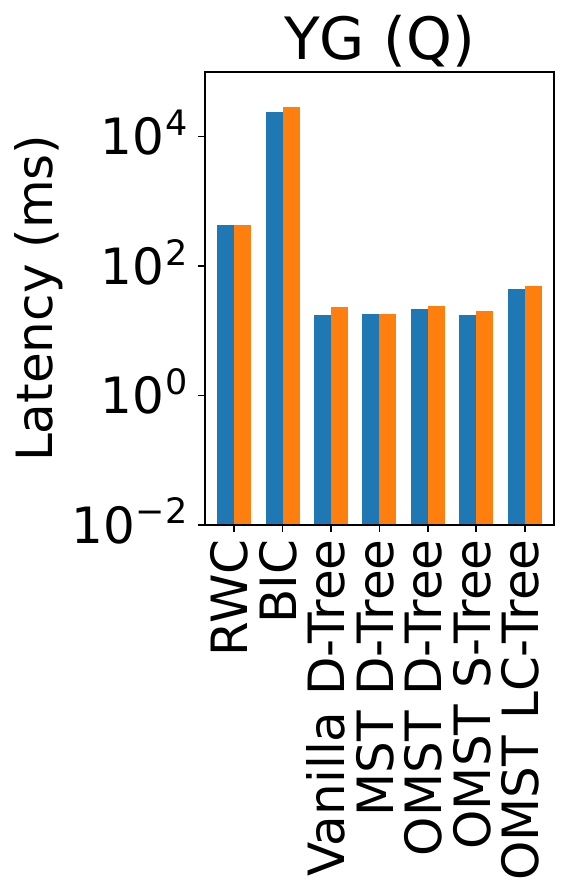}
    }
\end{minipage}
\begin{minipage}{0.094\linewidth}
\resizebox{\textwidth}{!}{
    \includegraphics{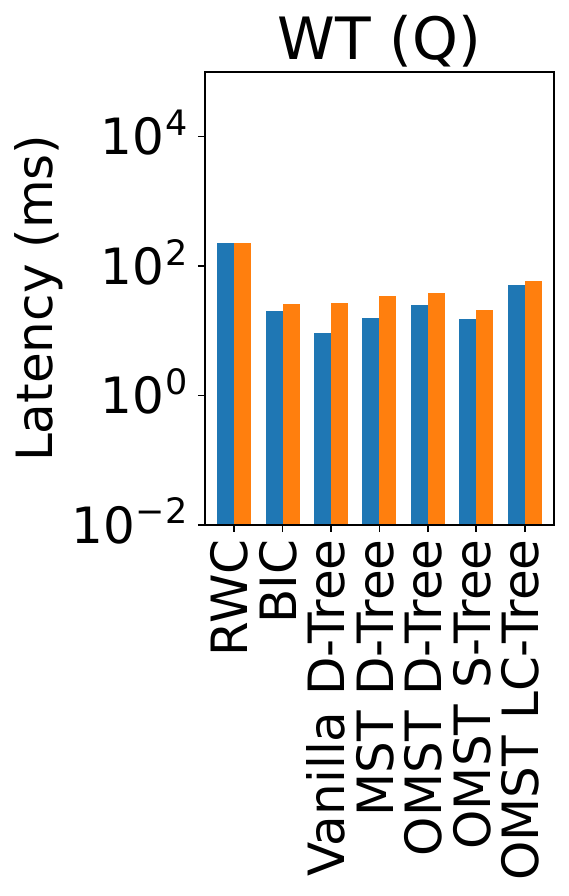}
    }
\end{minipage}
\begin{minipage}{0.094\linewidth}
\resizebox{\textwidth}{!}{
    \includegraphics{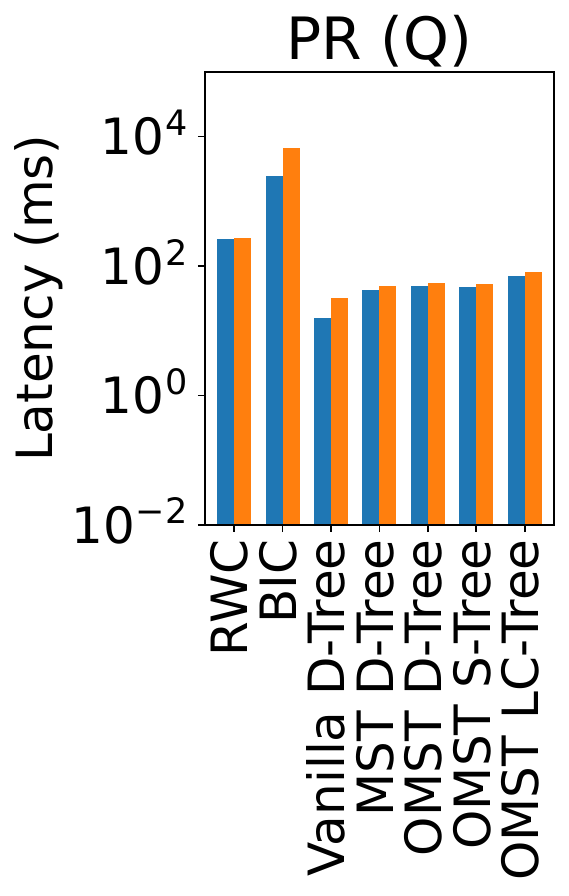}
    }
\end{minipage}
\begin{minipage}{0.094\linewidth}
\resizebox{\textwidth}{!}{
    \includegraphics{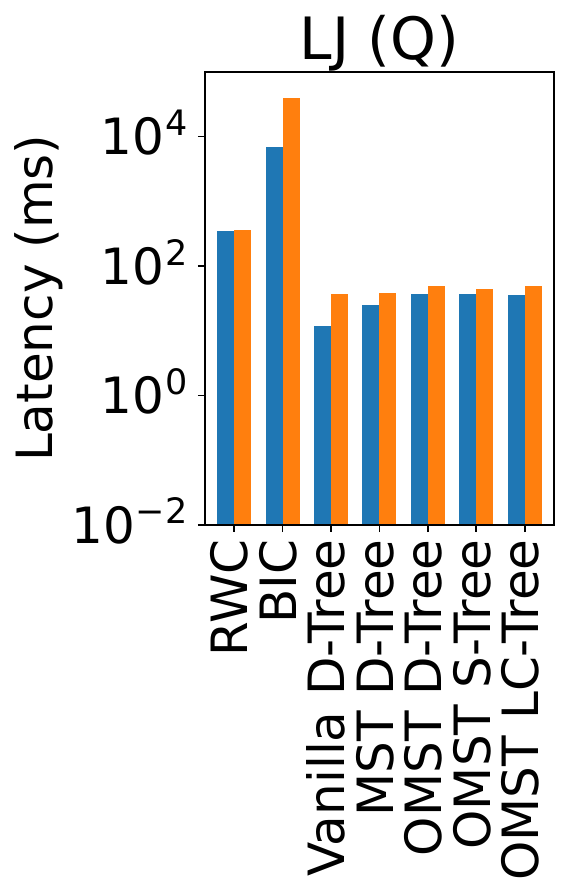}
    }
\end{minipage}
\begin{minipage}{0.094\linewidth}
\resizebox{\textwidth}{!}{
    \includegraphics{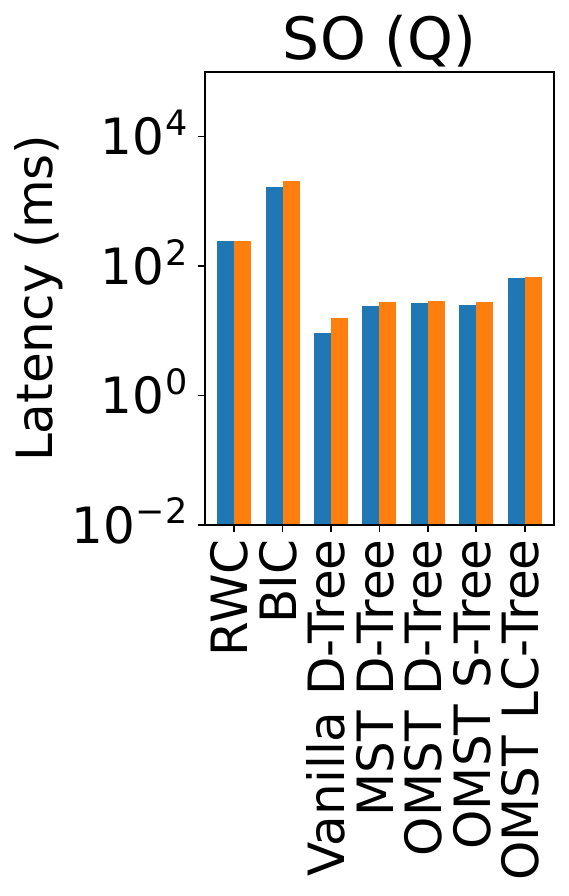}
    }
\end{minipage}
\begin{minipage}{0.094\linewidth}
\resizebox{\textwidth}{!}{
    \includegraphics{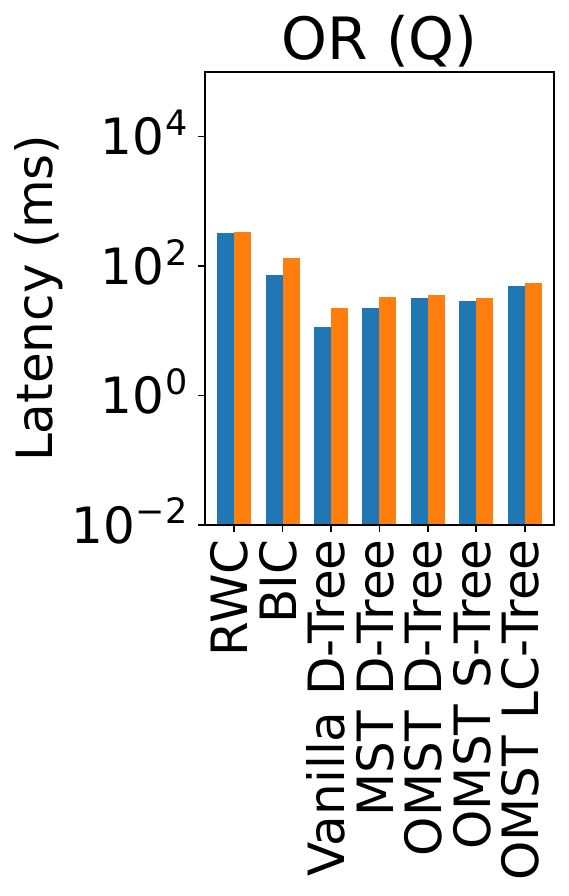}
    }
\end{minipage}
\begin{minipage}{0.094\linewidth}
\resizebox{\textwidth}{!}{
    \includegraphics{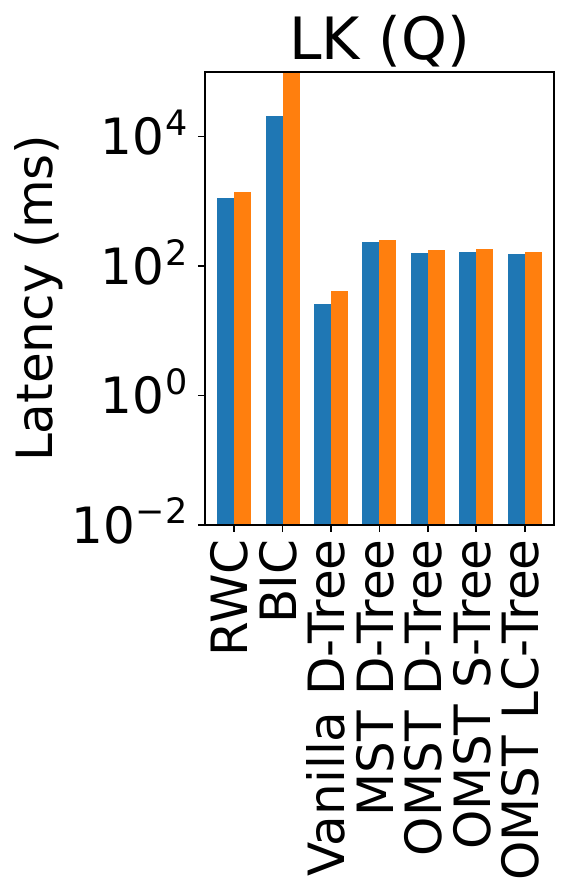}
    }
\end{minipage}
\begin{minipage}{0.094\linewidth}
\resizebox{\textwidth}{!}{
    \includegraphics{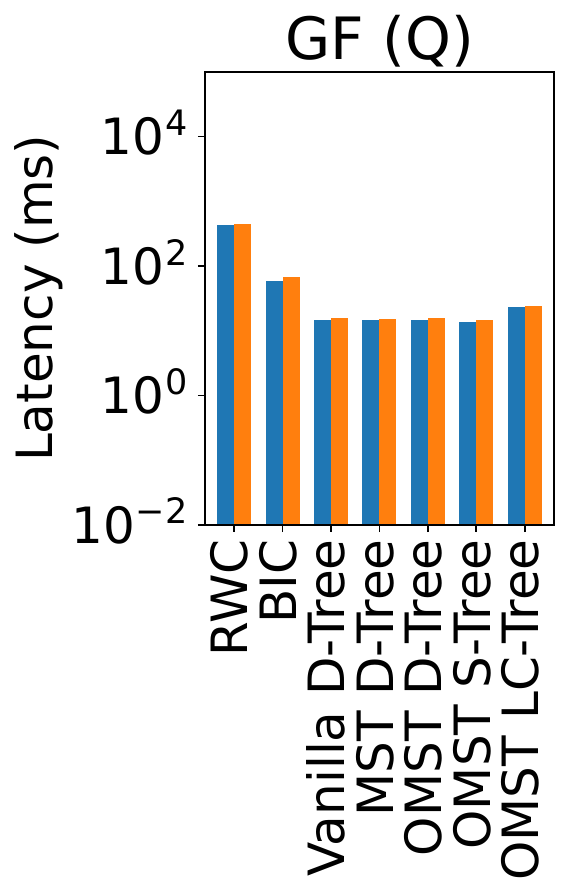}
    }
\end{minipage}
\begin{minipage}{0.094\linewidth}
\resizebox{\textwidth}{!}{
    \includegraphics{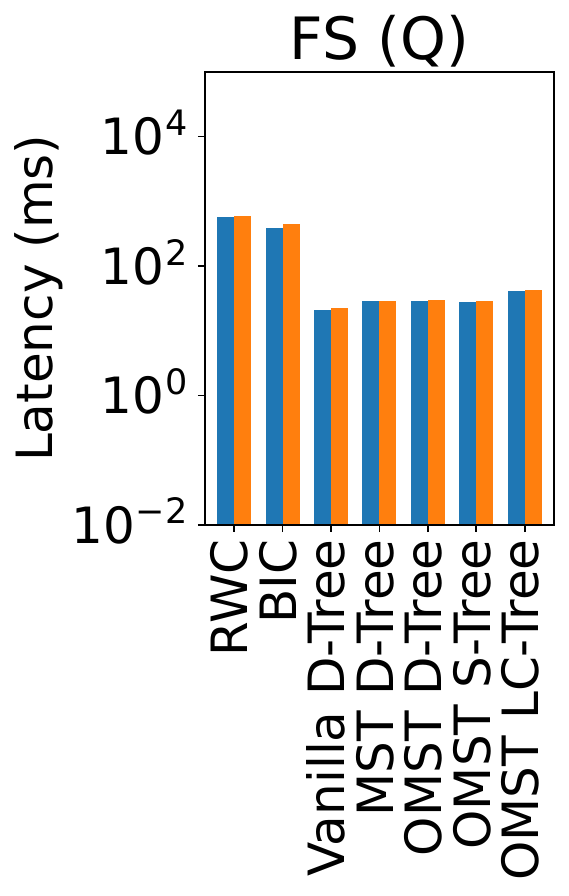}
    }
\end{minipage}
\begin{minipage}{0.094\linewidth}
\resizebox{\textwidth}{!}{
    \includegraphics{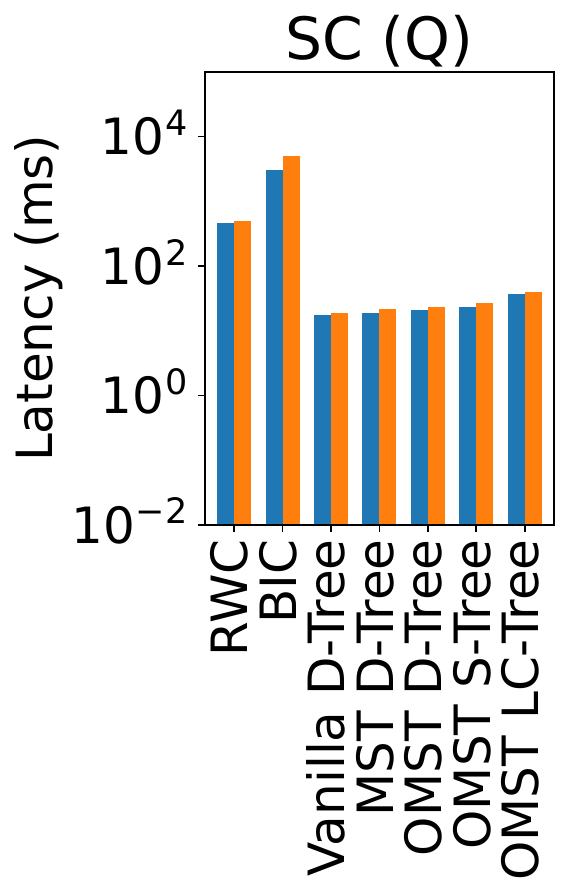}
    }
\end{minipage}
\begin{minipage}{0.094\linewidth}
\resizebox{\textwidth}{!}{
    \includegraphics{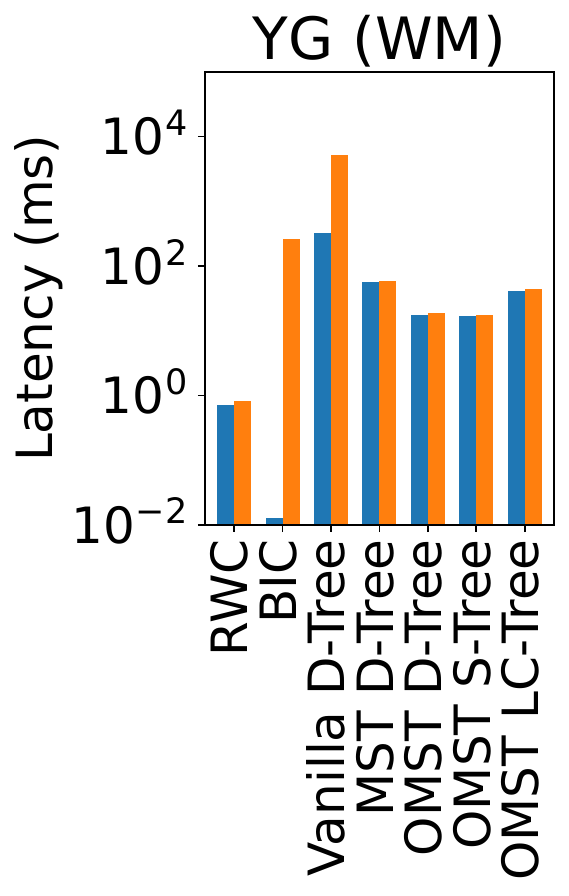}
    }
\end{minipage}
\begin{minipage}{0.094\linewidth}
\resizebox{\textwidth}{!}{
    \includegraphics{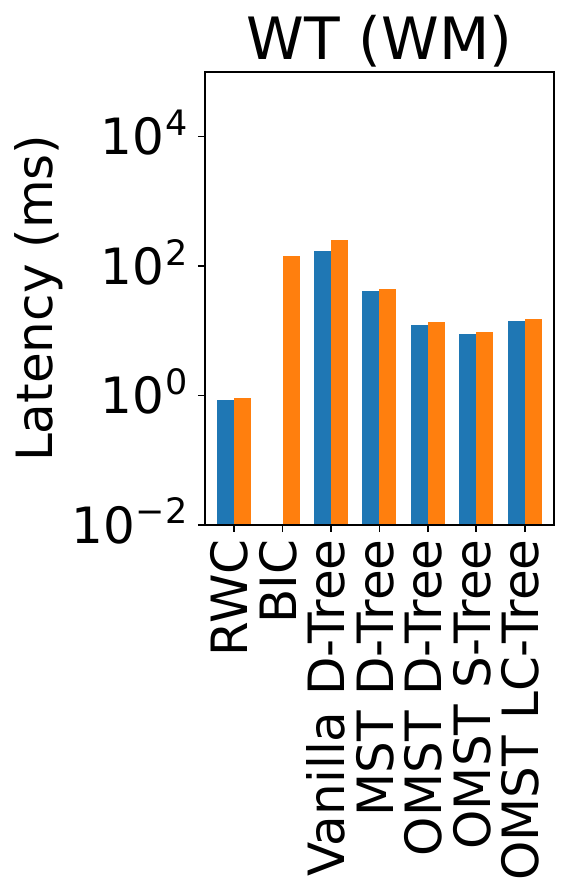}
    }
\end{minipage}
\begin{minipage}{0.094\linewidth}
\resizebox{\textwidth}{!}{
    \includegraphics{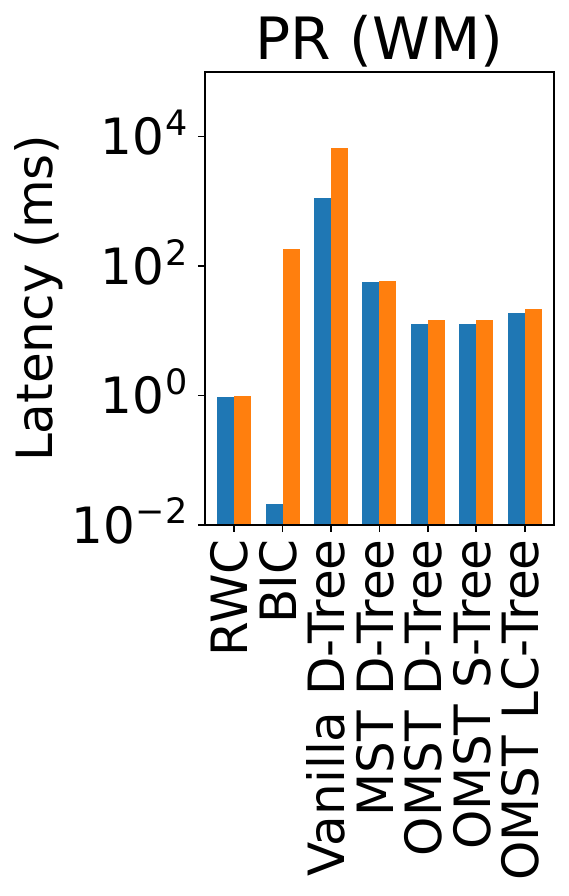}
    }
\end{minipage}
\begin{minipage}{0.094\linewidth}
\resizebox{\textwidth}{!}{
    \includegraphics{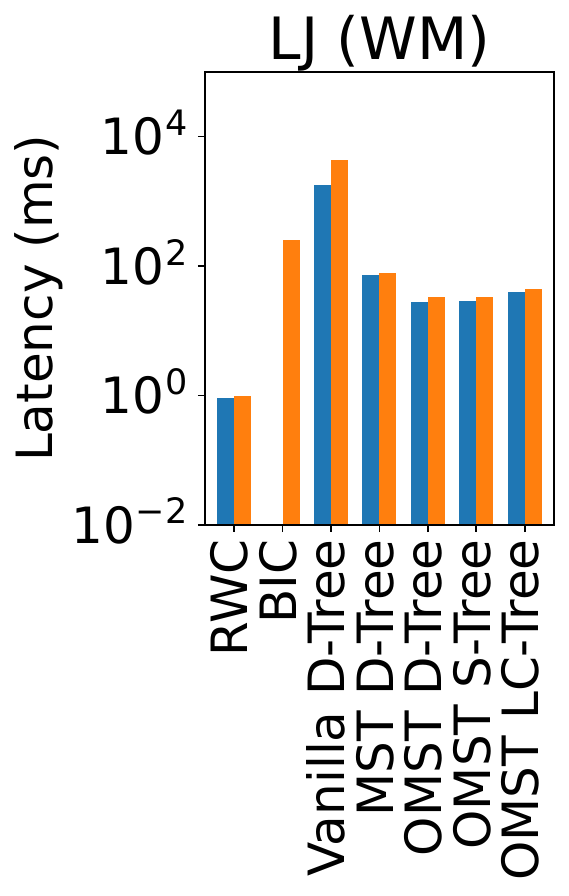}
    }
\end{minipage}
\begin{minipage}{0.094\linewidth}
\resizebox{\textwidth}{!}{
    \includegraphics{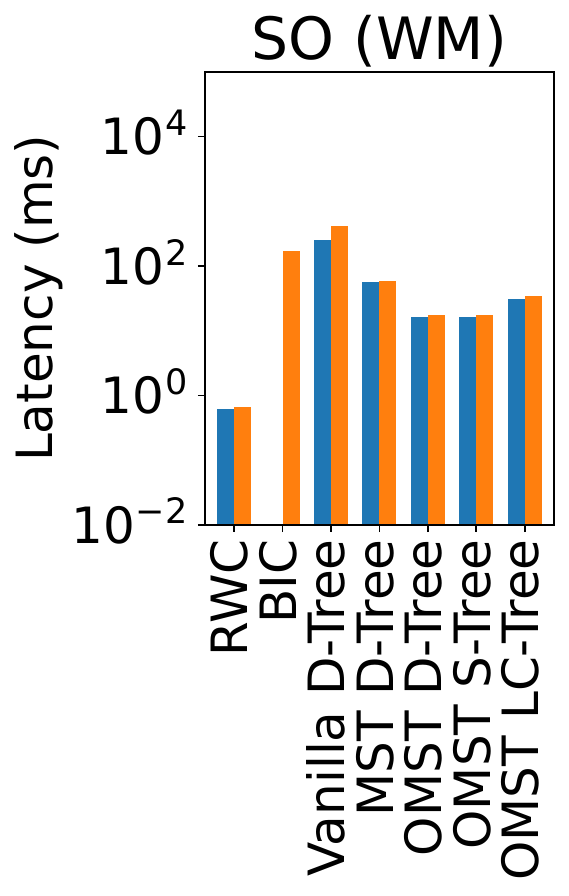}
    }
\end{minipage}
\begin{minipage}{0.094\linewidth}
\resizebox{\textwidth}{!}{
    \includegraphics{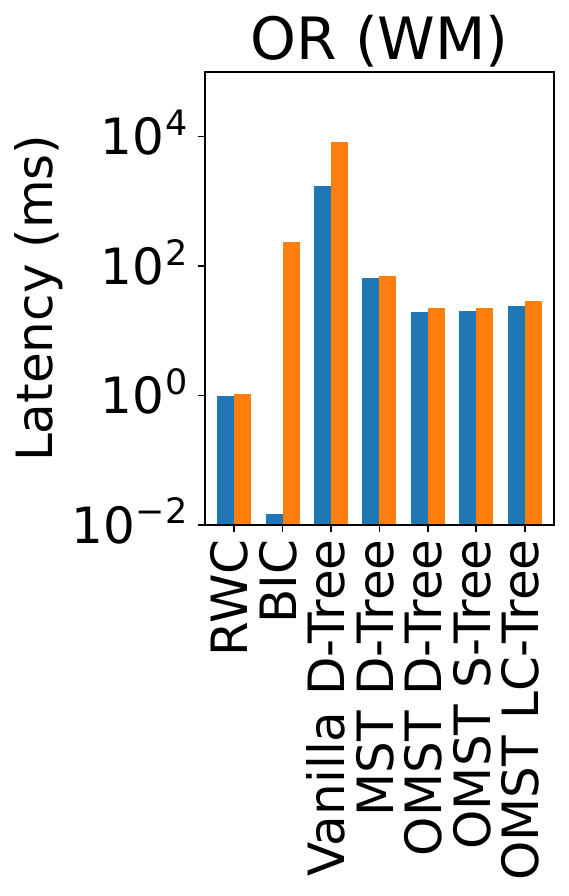}
    }
\end{minipage}
\begin{minipage}{0.094\linewidth}
\resizebox{\textwidth}{!}{
    \includegraphics{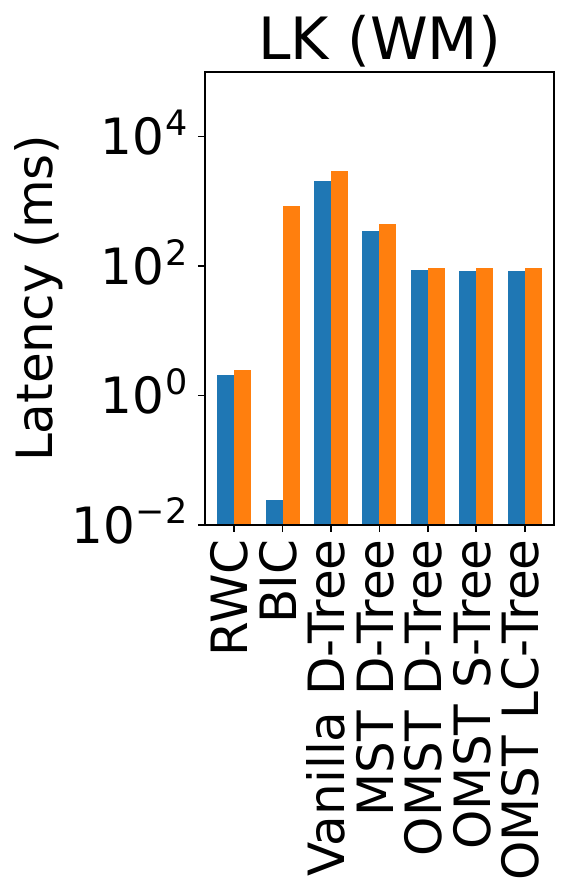}
    }
\end{minipage}
\begin{minipage}{0.094\linewidth}
\resizebox{\textwidth}{!}{
    \includegraphics{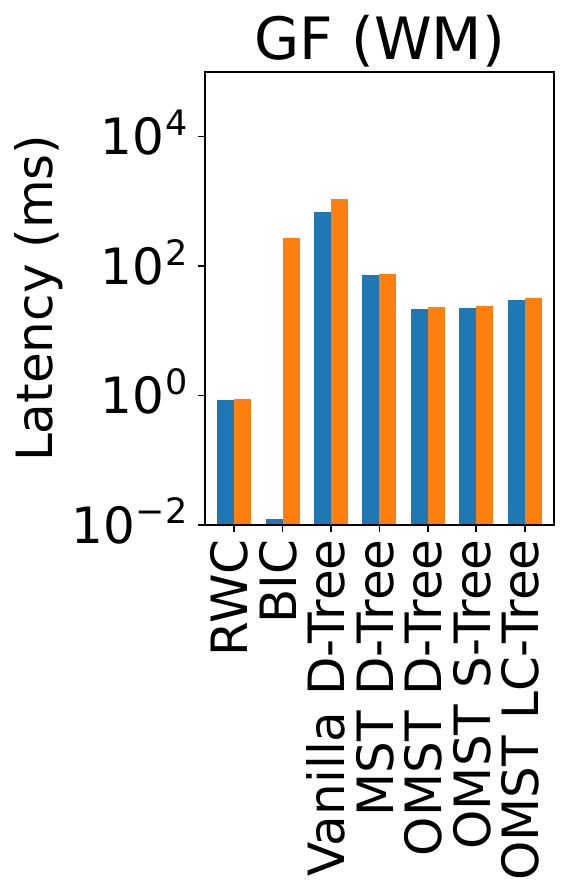}
    }
\end{minipage}
\begin{minipage}{0.094\linewidth}
\resizebox{\textwidth}{!}{
    \includegraphics{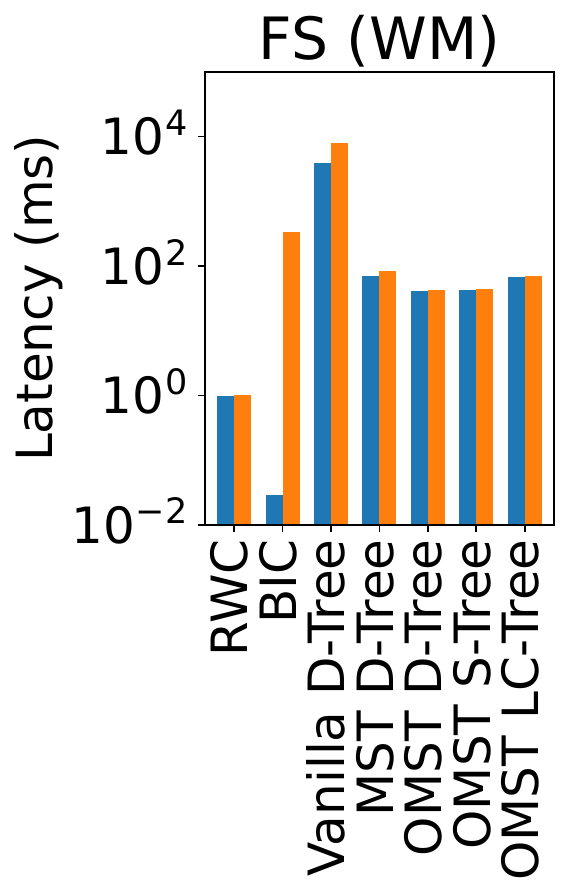}
    }
\end{minipage}
\begin{minipage}{0.094\linewidth}
\resizebox{\textwidth}{!}{
    \includegraphics{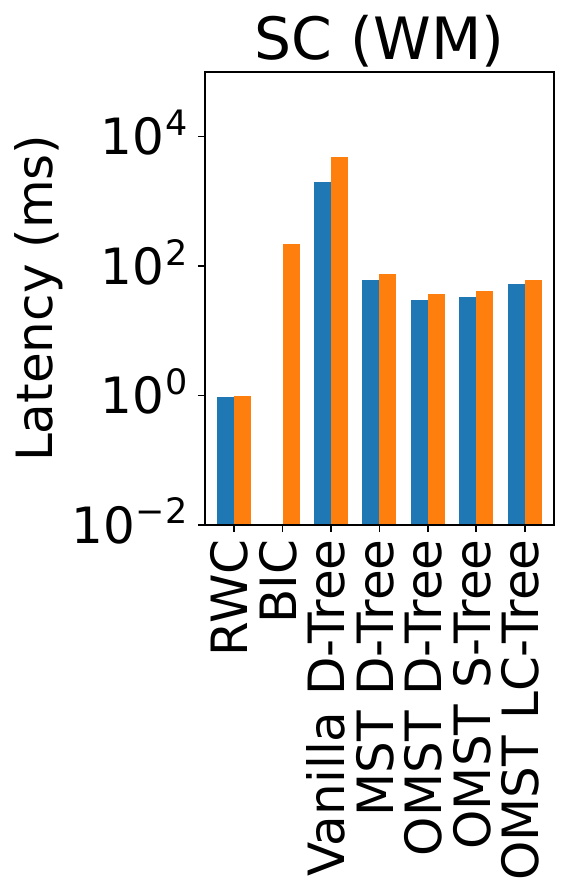}
    }
\end{minipage}
\caption{Query (Q) latency and window management (WM) latency results.}\label{fig:per_latency}
\end{figure*}

\begin{figure*}
\begin{minipage}{0.095\linewidth}
\resizebox{\textwidth}{!}{
    \includegraphics{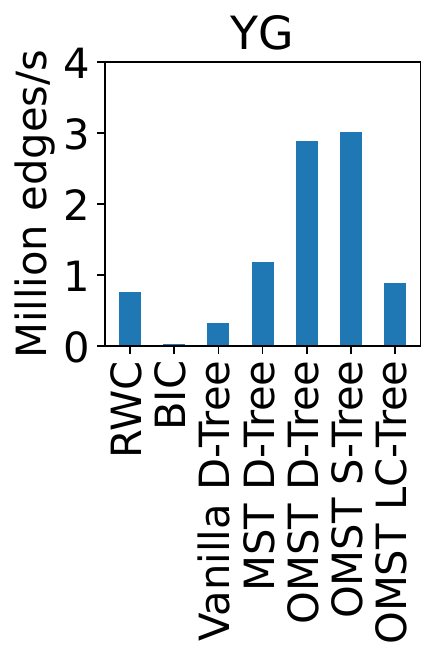}
    }    
\end{minipage}
\begin{minipage}{0.095\linewidth}
\resizebox{\textwidth}{!}{
    \includegraphics{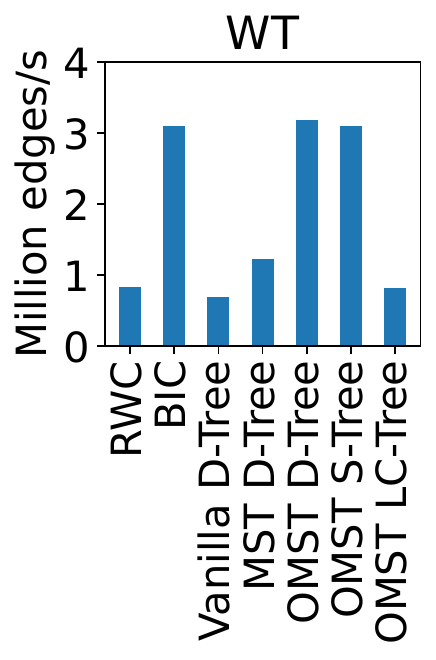}
    }
\end{minipage}
\begin{minipage}{0.095\linewidth}
\resizebox{\textwidth}{!}{
    \includegraphics{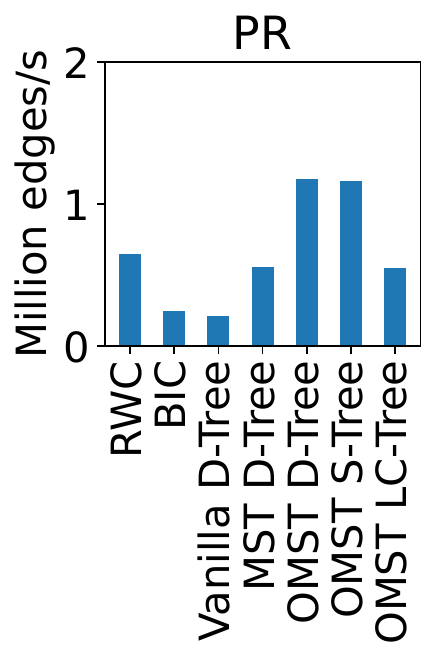}
    }
\end{minipage}
\begin{minipage}{0.095\linewidth}
\resizebox{\textwidth}{!}{
    \includegraphics{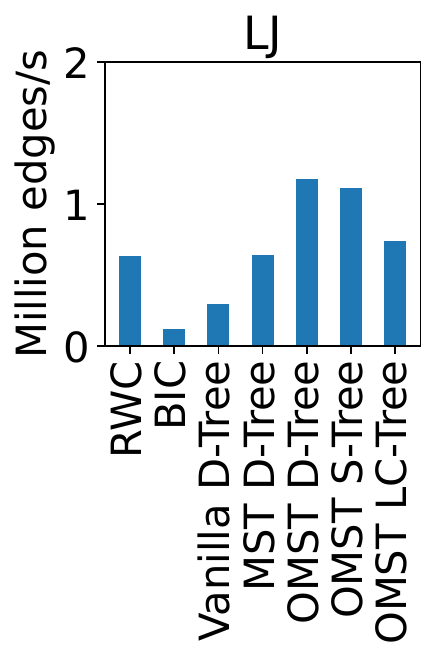}
    }
\end{minipage}
\begin{minipage}{0.095\linewidth}
\resizebox{\textwidth}{!}{
    \includegraphics{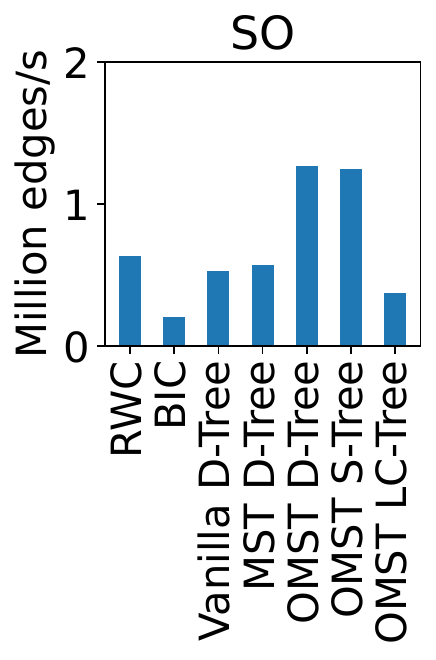}
    }
\end{minipage}
\begin{minipage}{0.095\linewidth}
\resizebox{\textwidth}{!}{
    \includegraphics{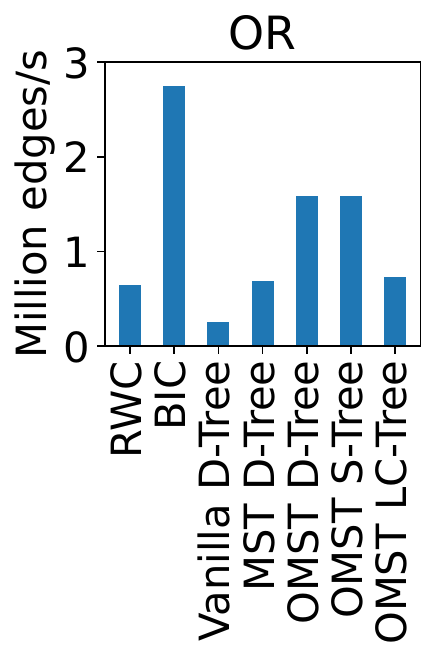}
    }
\end{minipage}
\begin{minipage}{0.095\linewidth}
\resizebox{\textwidth}{!}{
    \includegraphics{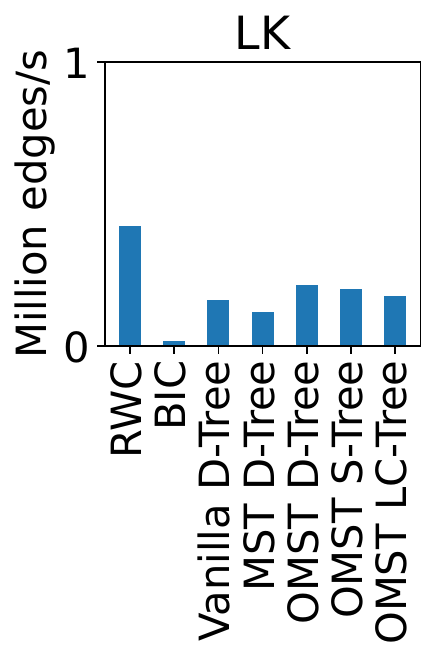}
    }
\end{minipage}
\begin{minipage}{0.095\linewidth}
\resizebox{\textwidth}{!}{
    \includegraphics{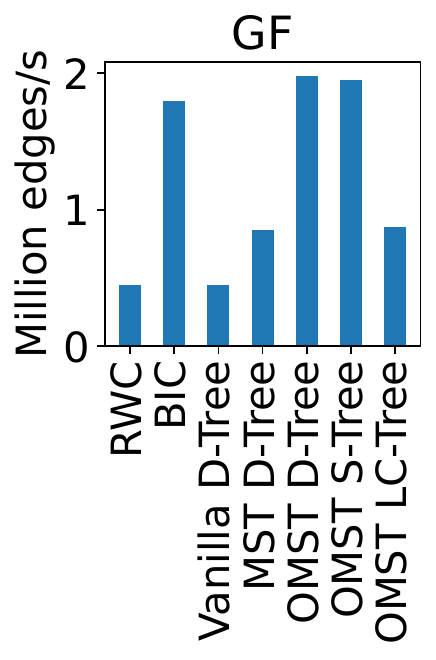}
    }
\end{minipage}
\begin{minipage}{0.095\linewidth}
\resizebox{\textwidth}{!}{
    \includegraphics{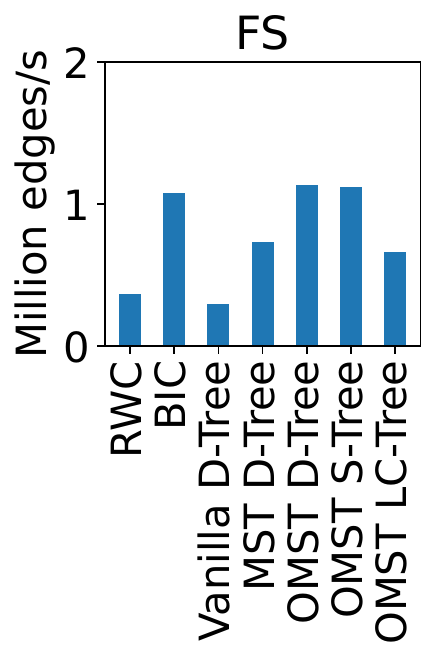}
    }
\end{minipage}
\begin{minipage}{0.095\linewidth}
\resizebox{\textwidth}{!}{
    \includegraphics{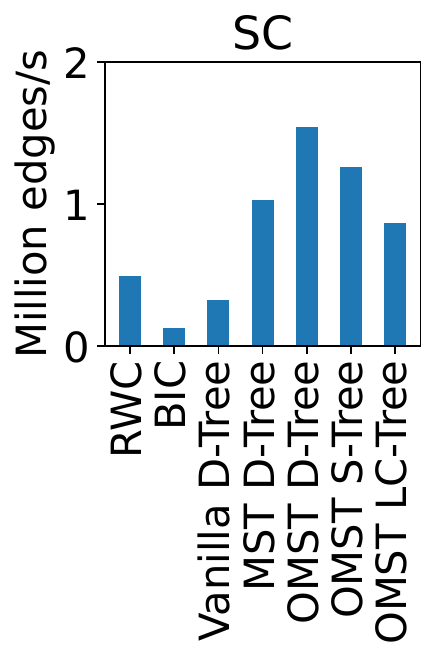}
    }
\end{minipage}
\caption{Throughput results.}\label{fig:per_throughput}
\end{figure*}

\subsection{Tail Latency and Throughput}\label{sec:performance}
Figure \ref{fig:per_latency} shows P95 and P99 latency for query and window management, while Figure \ref{fig:per_throughput} reports throughput. Both experiments use windows with an average size of $3$M edges, slide intervals averaging $150$K edges, and workloads of $10^5$ queries. 

\textit{Tail latency}. 
We present the results for query (Q) latency and window management (WM) latency in Figure \ref{fig:per_latency}. 
Generally, OMST D-Tree and OMST S-Tree, the most efficient among the proposed methods, demonstrate up to a $29\times$ reduction over RWC in query latency, up to a $1172\times$ and $13\times$ reduction over BIC in query and window management latency, respectively, and up to a $458\times$ reduction over Vanilla D-Tree in window management latency.
For query latency, indexing approaches based on spanning trees—including Vanilla D-Tree, MST D-Tree, OMST S-Tree, OMST D-Tree, and OMST LC-Tree—exhibit similar performance and significantly outperform RWC and BIC. 
RWC incurs high query latency as it requires computing all connected components within the window before processing queries. BIC also suffers from high query latency for the tested workload of $10^5$ queries, due to the need to traverse a bipartite graph to determine query results in the worst case. In contrast, spanning tree-based approaches can compute query results by simply traversing the path from vertices to their roots. 
For window management latency, RWC does not maintain an index, and its latency only involves the deletion of expired edges from the window. Among index-based approaches, our methods—MST D-Tree, OMST D-Tree, OMST S-Tree, and OMST LC-Tree—significantly outperform the others. BIC experiences high P99 window management latency due to its backward computation over an entire chunk, though this costly operation occurs only once every $\alpha / \beta$ window, resulting in a relatively low P95 window management latency.
Vanilla D-Tree exhibits notably high P95 and P99 window management latency because deleting expired edges from its spanning tree requires searching for replacement edges in the worst case. Our methods show the best P99 latency results, as deleting expired edges from spanning trees in our framework avoids the need to search for replacement edges. The significant reduction in window management latency from Vanilla D-Tree to MST D-Tree, and further to OMST D-Tree, underscores the effectiveness of the MST framework and the optimizations introduced by the OMST framework.
OMST S-Tree demonstrates comparable latency results to OMST D-Tree, indicating that our proposed framework can achieve exceptionally low latency even with the integration of simple techniques. 
OMST LC-Tree, the most theoretically intriguing approach, demonstrates comparably low latency on both LJ and LK. However, it depends on a more complex \texttt{access} operation for updating spanning trees and processing queries, leading to slightly higher latency compared to OMST D-Tree and OMST S-Tree.

\textit{Throughput}.
The throughput results are presented in Figure \ref{fig:per_throughput}. 
Generally, OMST D-Tree and OMST S-Tree, which exhibit the best latency results, also outperform existing methods in terms of throughput, demonstrating up to $80\times$ improvement over BIC, $4\times$ over RWC, and $8\times$ over Vanilla D-Tree. These results highlight their capability to achieve high throughput and low latency in computing sliding window connectivity.
There are a few exceptions to note; BIC performs well the OR graph, while RWC excels on the LK graph. 
BIC, operating as a partial index, computes full connectivity information only during query processing, allowing high throughput for smaller workloads but facing scalability issues. 
Conversely, RWC computes connected components after the window completes, distributing costs across all edges but resulting in higher latency.
Although the OMST framework involves minimal overhead in handling expired edges, it requires specific computations to maintain maximum spanning trees within sliding windows. This requirement accounts for the lesser throughput improvement of OMST D-Tree and OMST S-Tree compared to their more significant latency improvements over other methods.
Nevertheless, OMST D-Tree and OMST S-Tree consistently outperform existing methods across most graphs, particularly in real-world datasets.
The sequential improvements from Vanilla D-Tree to MST D-Tree and then to OMST D-Tree highlight the effectiveness of the MST and OMST frameworks. 
Additionally, OMST S-Tree shows performance comparable to OMST D-Tree, demonstrating the robustness of the OMST framework, even with simpler techniques.
However, OMST LC-Tree's throughput is lower than OMST S-Tree and OMST D-Tree due to the complex \texttt{access} operations required for inserting each streaming edge, which must be executed multiple times.

\begin{figure*}
\centering
\resizebox{0.88\linewidth}{!}{
\includegraphics{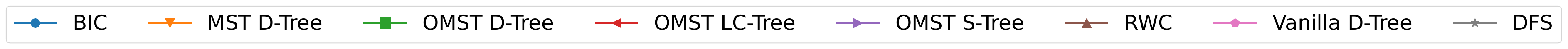}}\\
\begin{minipage}{0.29\linewidth}
    \begin{minipage}{0.32\textwidth}
        \resizebox{\textwidth}{!}{
\includegraphics{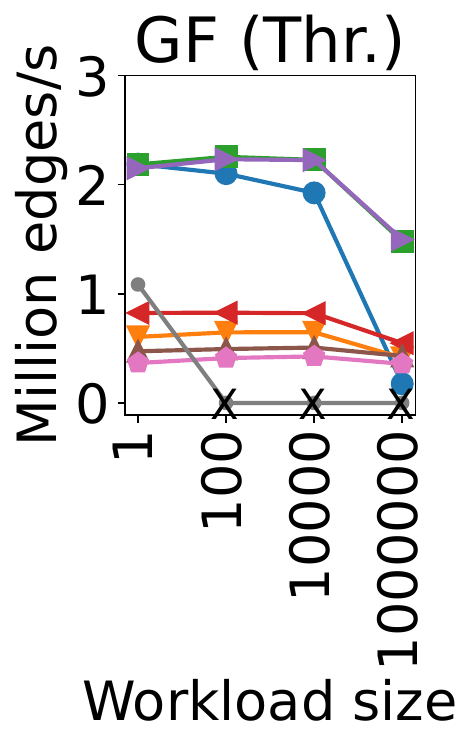}
}
    \end{minipage}
    \begin{minipage}{0.32\textwidth}
        \resizebox{\textwidth}{!}{
\includegraphics{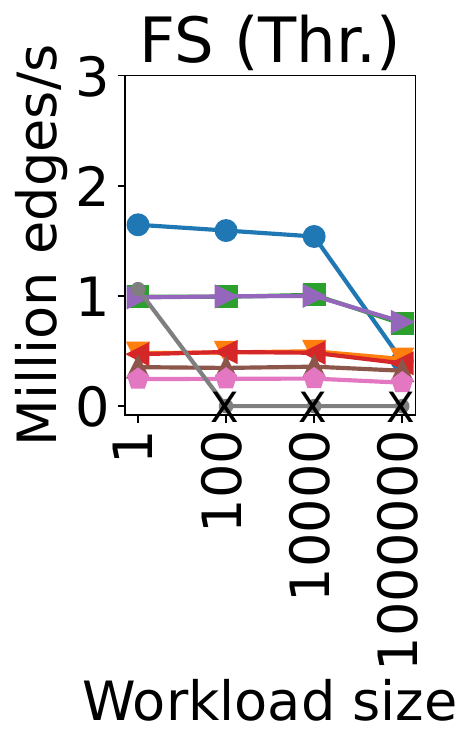}
}
    \end{minipage}
    \begin{minipage}{0.32\textwidth}
        \resizebox{\textwidth}{!}{
\includegraphics{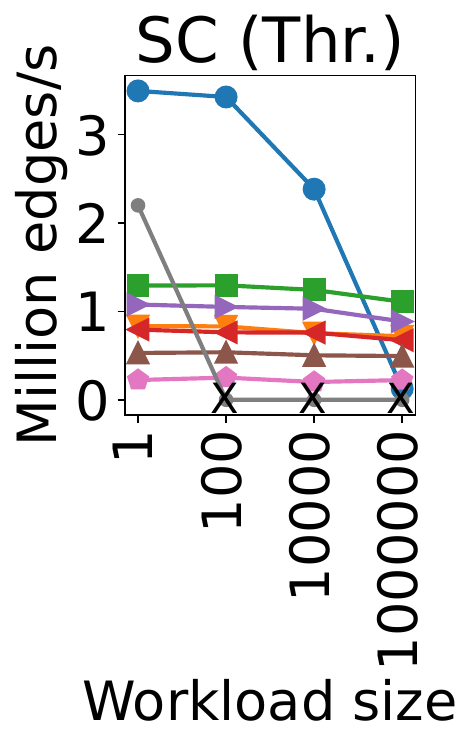}
}
    \end{minipage}
\end{minipage}
\begin{minipage}{0.345\linewidth}
    \begin{minipage}{0.32\textwidth}
        \resizebox{\textwidth}{!}{
\includegraphics{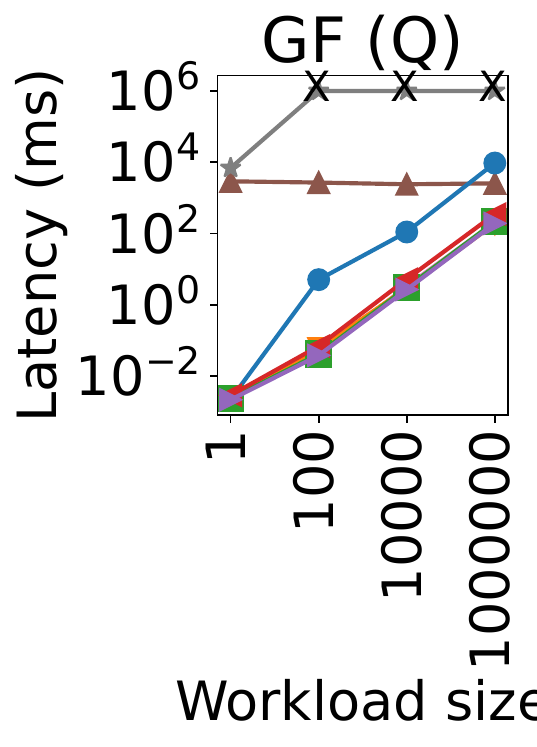}
}
    \end{minipage}
    \begin{minipage}{0.32\textwidth}
        \resizebox{\textwidth}{!}{
\includegraphics{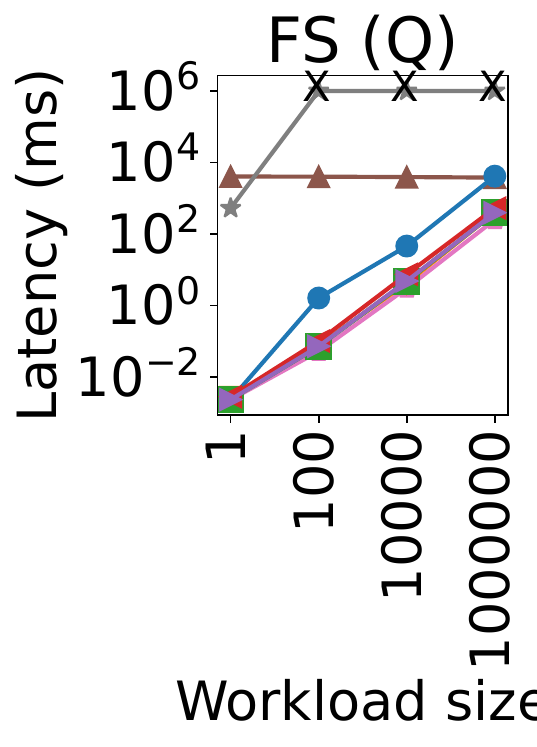}
}
    \end{minipage}
    \begin{minipage}{0.32\textwidth}
        \resizebox{\textwidth}{!}{
\includegraphics{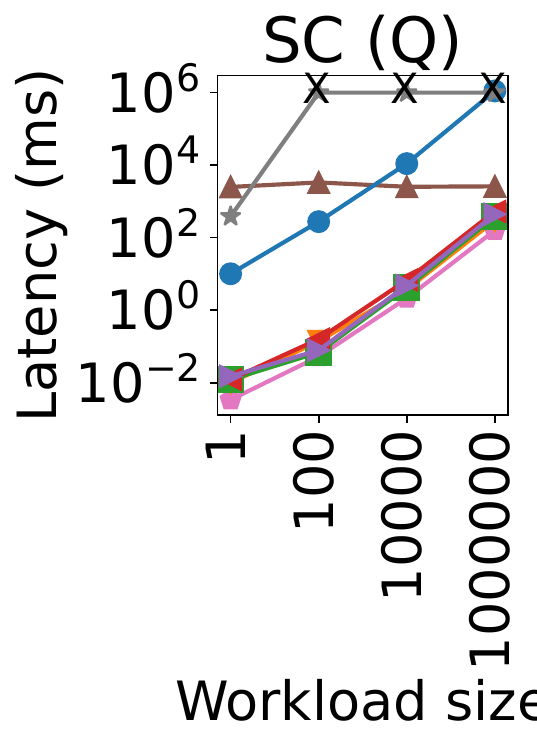}
}
    \end{minipage}
\end{minipage}
\begin{minipage}{0.345\linewidth}
    \begin{minipage}{0.32\textwidth}
        \resizebox{\textwidth}{!}{
\includegraphics{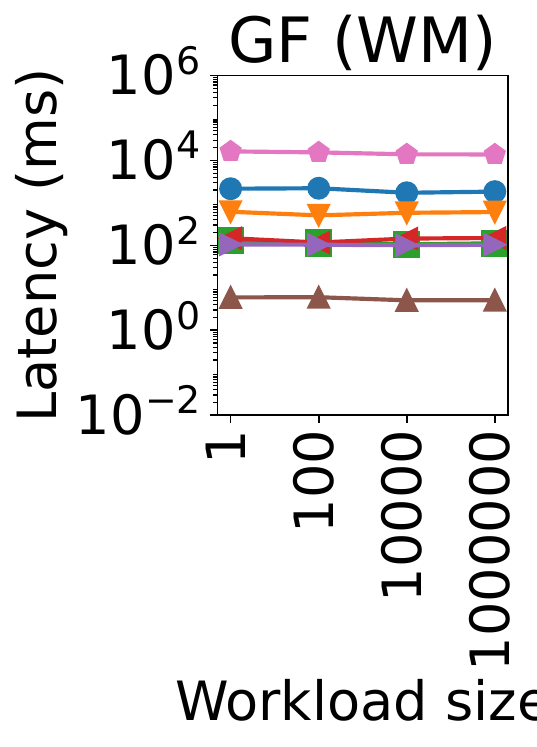}
}
    \end{minipage}
    \begin{minipage}{0.32\textwidth}
        \resizebox{\textwidth}{!}{
\includegraphics{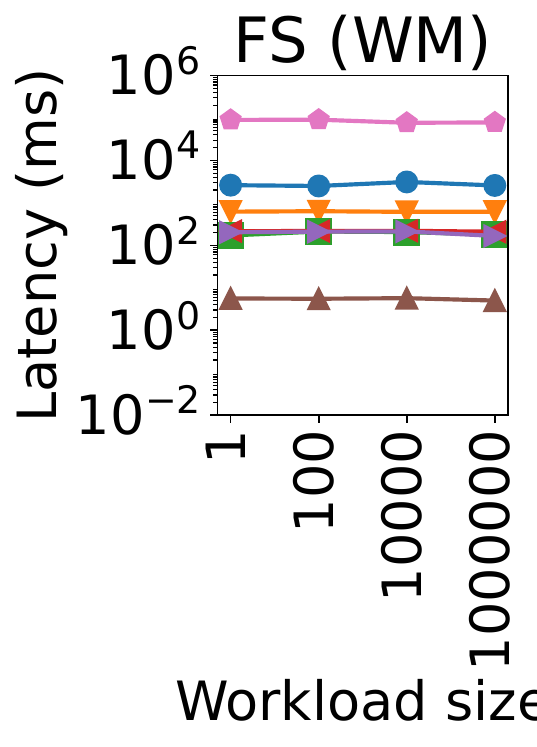}
}
    \end{minipage}
    \begin{minipage}{0.32\textwidth}
        \resizebox{\textwidth}{!}{
\includegraphics{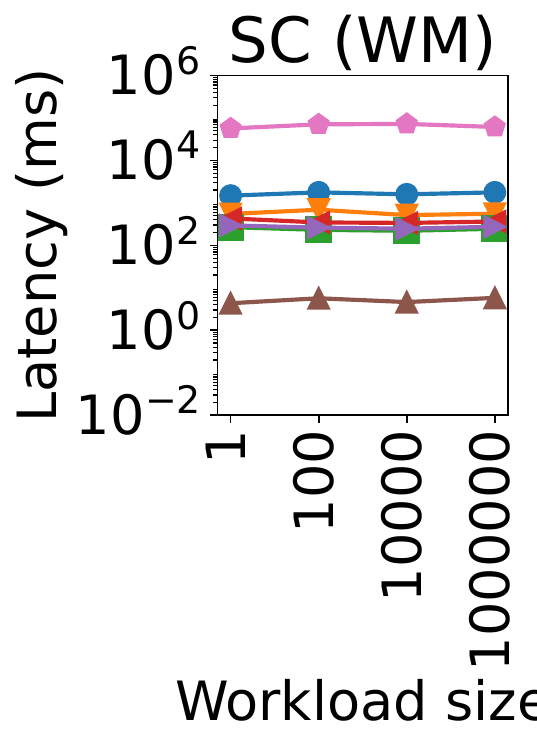}
}
    \end{minipage}
\end{minipage}
\caption{Throughput, and P99 latency for query (Q) and window management (WM) using workloads of various sizes.}\label{fig:scale_workload_size}
\end{figure*}

\begin{figure*}
\centering
\resizebox{0.8\linewidth}{!}{
\includegraphics{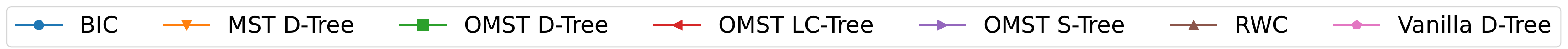}}\\
\begin{minipage}{0.29\linewidth}
    \begin{minipage}{0.32\textwidth}
        \resizebox{\textwidth}{!}{
            \includegraphics{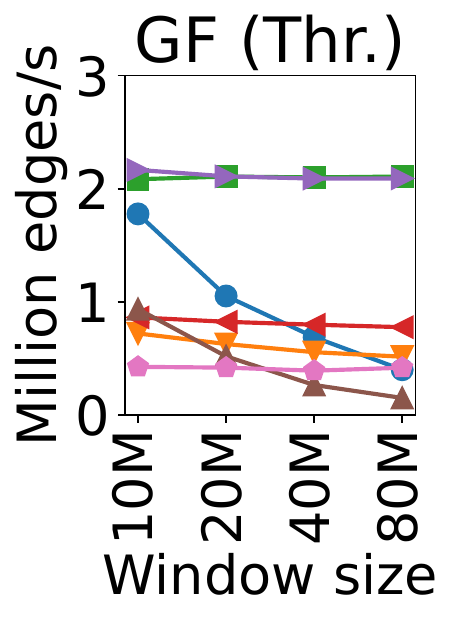}
        }
    \end{minipage}
    \begin{minipage}{0.32\textwidth}
        \resizebox{\textwidth}{!}{
            \includegraphics{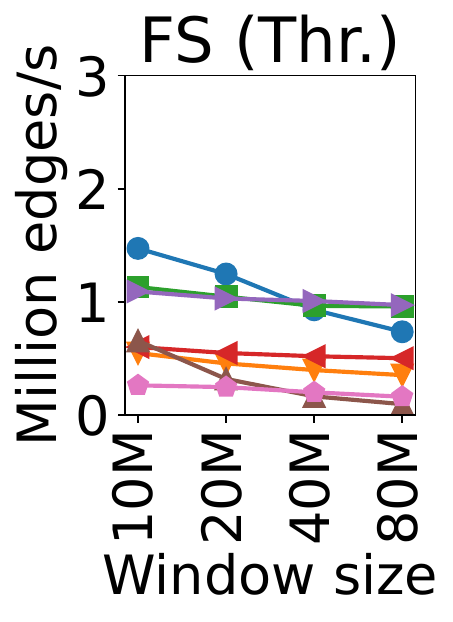}
        }
    \end{minipage}
    \begin{minipage}{0.32\textwidth}
        \resizebox{\textwidth}{!}{
            \includegraphics{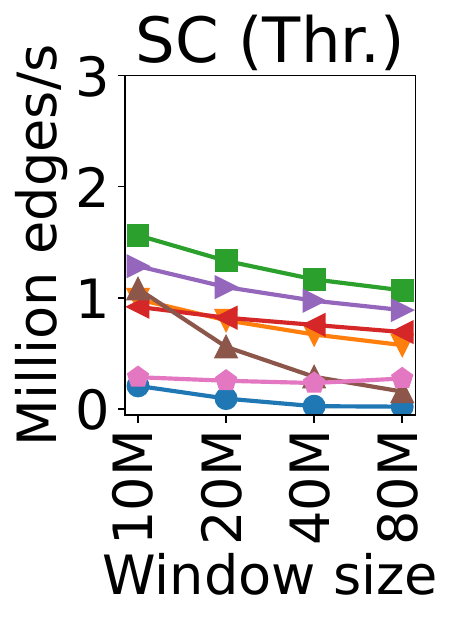}
        }
    \end{minipage}
\end{minipage}
\begin{minipage}{0.345\linewidth}
    \begin{minipage}{0.32\textwidth}
        \resizebox{\textwidth}{!}{
\includegraphics{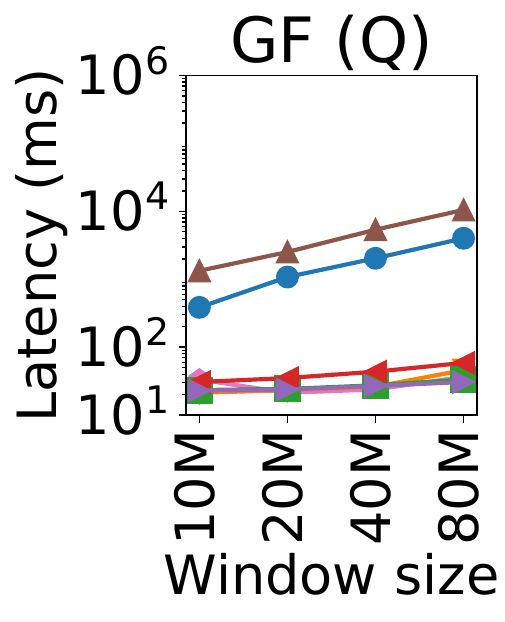}
}
    \end{minipage}
    \begin{minipage}{0.32\textwidth}
        \resizebox{\textwidth}{!}{
\includegraphics{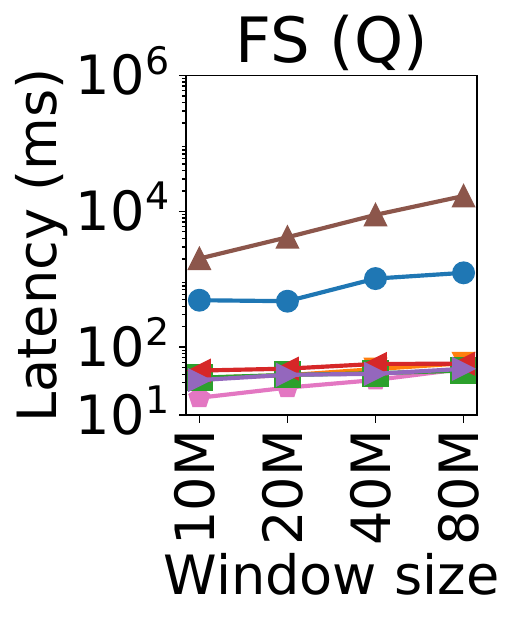}
}
    \end{minipage}
    \begin{minipage}{0.32\textwidth}
        \resizebox{\textwidth}{!}{
\includegraphics{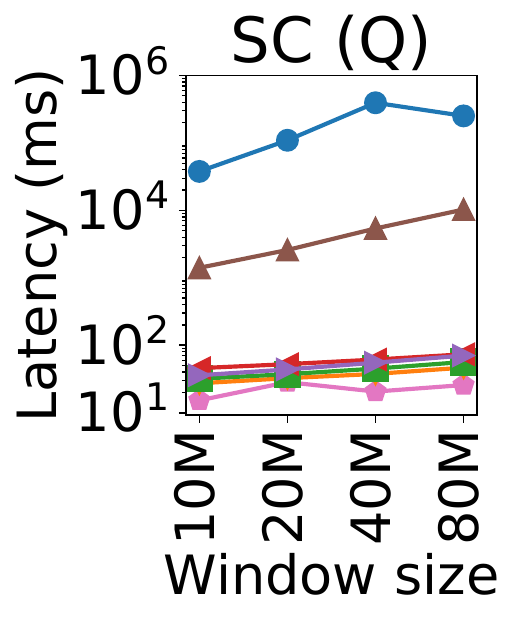}
}
    \end{minipage}
\end{minipage}
\begin{minipage}{0.345\linewidth}
    \begin{minipage}{0.32\textwidth}
        \resizebox{\textwidth}{!}{
\includegraphics{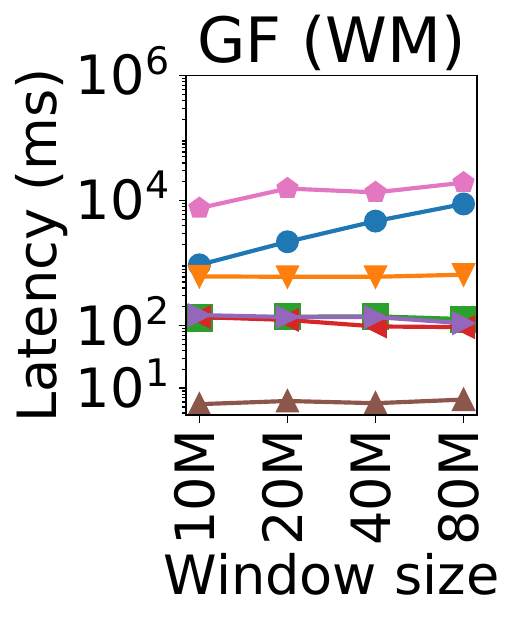}
}
    \end{minipage}
    \begin{minipage}{0.32\textwidth}
        \resizebox{\textwidth}{!}{
\includegraphics{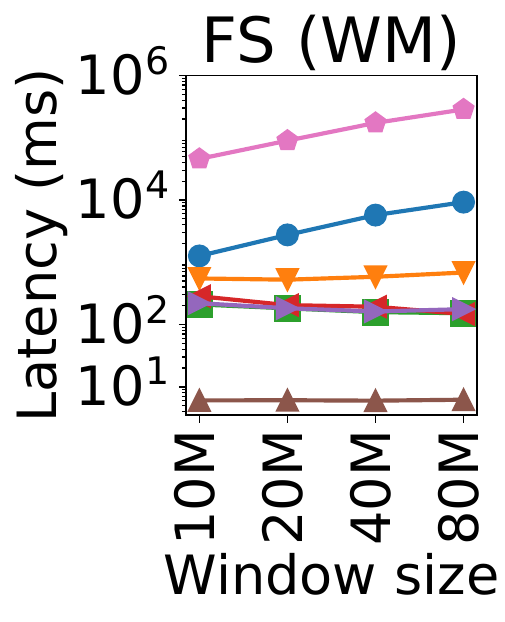}
}
    \end{minipage}
        \begin{minipage}{0.32\textwidth}
        \resizebox{\textwidth}{!}{
\includegraphics{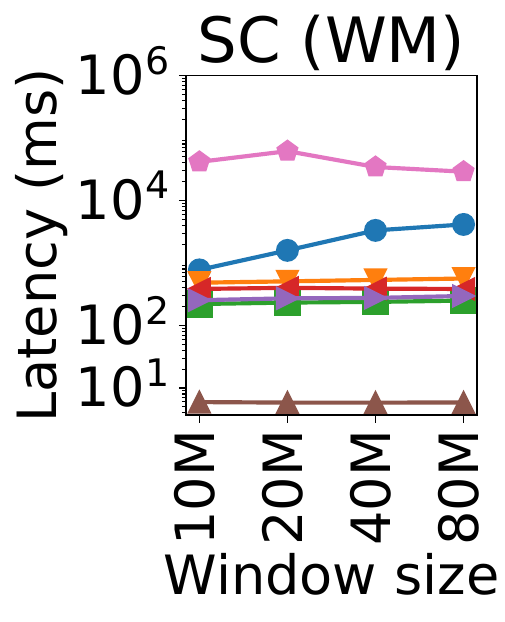}
}
    \end{minipage}    
\end{minipage}
\caption{Throughput, and P99 latency for query (Q) and window management (WM) using windows of various sizes.}\label{fig:scale_window_size}
\end{figure*}

\subsection{Impact of Workload Sizes}\label{sec:workload}
Figure \ref{fig:scale_workload_size} presents the experimental results using a fixed window size of $20$M edges on average and a fixed slide interval of $1$M edges on average, while varying the workload sizes.
All methods, including the naive DFS approach, are evaluated. 
In general, all approaches experience a decline in throughput and an increase in query latency as the number of queries rises, due to the greater computational effort required for query processing. However, window management latency remains constant, as window sizes and slide intervals are unchanged. Our methods—MST D-Tree, OMST D-Tree, OMST S-Tree, and OMST LC-Tree—consistently outperform other methods with larger workloads.
%
The performance of DFS decreases significantly with increasing workload size due to repeated graph traversals for each query. This is further exacerbated by a notable increase in query latency. In the cases of large workload sizes, DFS encounters timeouts within a 10-hour limit, as indicated by `X' in the figures. This underscores the necessity for more efficient, index-based approaches.
%
The impact of workload size on RWC's throughput and window management latency is negligible, as the main performance bottleneck for RWC lies in computing connected components from scratch within sliding windows, which are then used for computing queries. 
As a result, RWC's query latency can be orders of magnitude higher than that of indexing approaches.
%
BIC's query latency increases with larger workload sizes and is significantly higher than that of spanning tree-based indexing approaches. This is because BIC, operating as a partial index, does not maintain complete connectivity information prior to query processing. As a result, it must perform extensive traversals across its backward and forward UFTs to evaluate global connectivity within the window, leading to considerable computational overhead.
Due to this high query latency, BIC experiences a substantial drop in throughput, particularly as the workload size increases from $10^4$ to $10^6$. It is important to note, however, that BIC can exhibit high throughput with smaller workloads, as its average window management latency remains low. This is because the most expensive operation—backward computation over a chunk—is only performed once every $\alpha /\beta$ windows.
%
For spanning tree-based approaches (Vanilla D-Tree, MST D-Tree, OMST D-Tree, and OMST S-Tree), although query latency increases with larger workloads, they consistently deliver the best query latency results due to their complete index structures.
Vanilla D-Tree, however, suffers from high window management latency due to the need to search for replacement edges when deleting expired edges, which significantly reduces its throughput compared to the proposed methods. Among these, OMST D-Tree and OMST S-Tree exhibit the most optimal performance across varying workload sizes.

\begin{figure*}
\centering
\resizebox{0.8\linewidth}{!}{
\includegraphics{figures/exp/scalability-throughput-legend.pdf}}\\
    \begin{minipage}{0.29\linewidth}
        \begin{minipage}{0.32\textwidth}
            \resizebox{\textwidth}{!}{
                \includegraphics{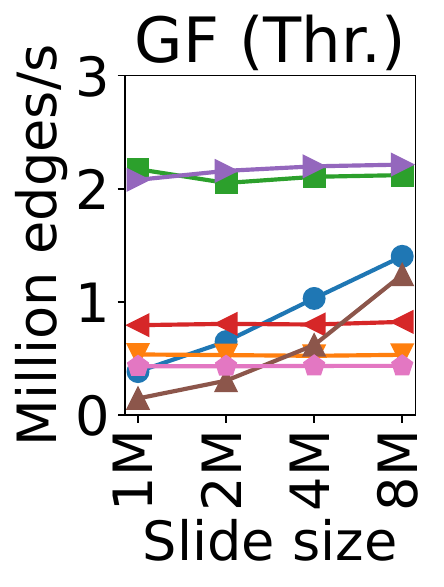}
            }
        \end{minipage}
        \begin{minipage}{0.32\textwidth}
            \resizebox{\textwidth}{!}{
                \includegraphics{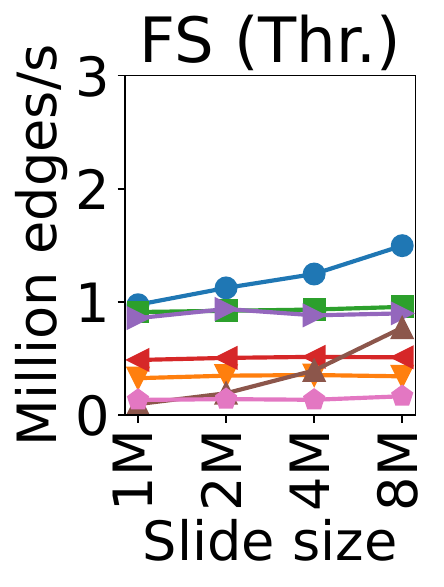}
            }
        \end{minipage}
        \begin{minipage}{0.32\textwidth}
            \resizebox{\textwidth}{!}{
                \includegraphics{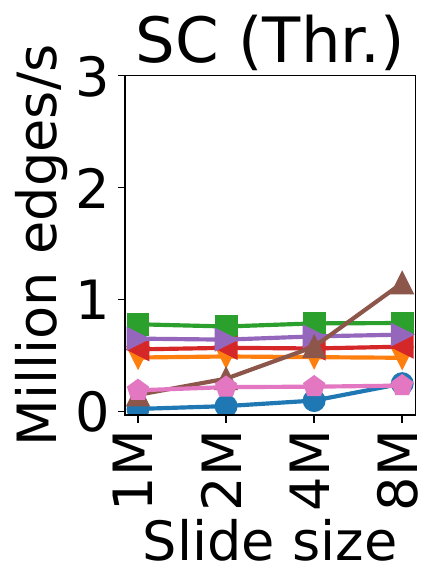}
            }
        \end{minipage}
    \end{minipage}
    \begin{minipage}{0.345\linewidth}
        \begin{minipage}{0.32\textwidth}
            \resizebox{\textwidth}{!}{
    \includegraphics{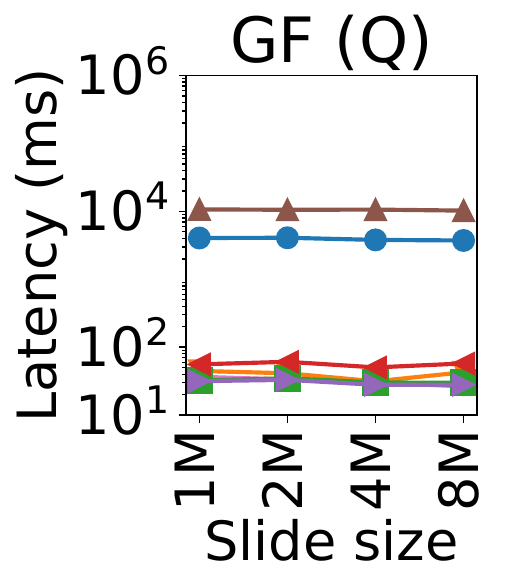}
    }
        \end{minipage}
        \begin{minipage}{0.32\textwidth}
            \resizebox{\textwidth}{!}{
    \includegraphics{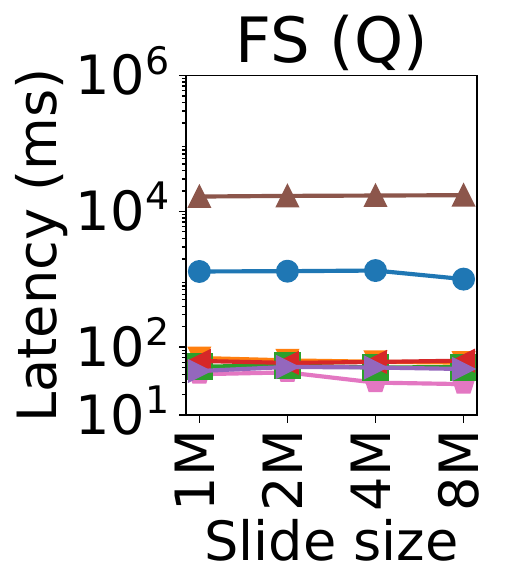}
    }
        \end{minipage}
        \begin{minipage}{0.32\textwidth}
            \resizebox{\textwidth}{!}{
    \includegraphics{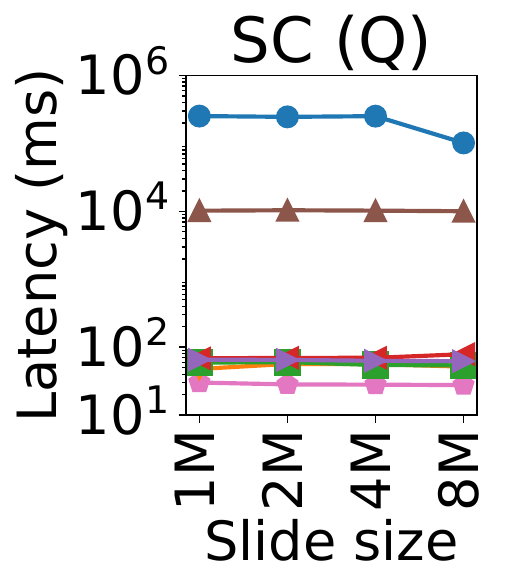}
    }
        \end{minipage}
    \end{minipage}
    \begin{minipage}{0.345\linewidth}
        \begin{minipage}{0.32\textwidth}
            \resizebox{\textwidth}{!}{
    \includegraphics{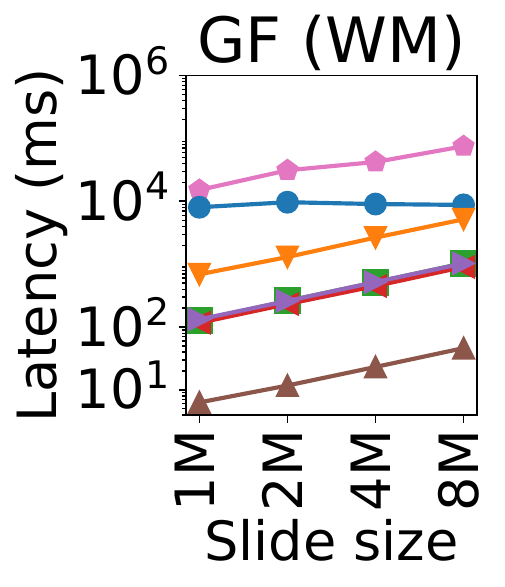}
    }
        \end{minipage}
        \begin{minipage}{0.32\textwidth}
            \resizebox{\textwidth}{!}{
    \includegraphics{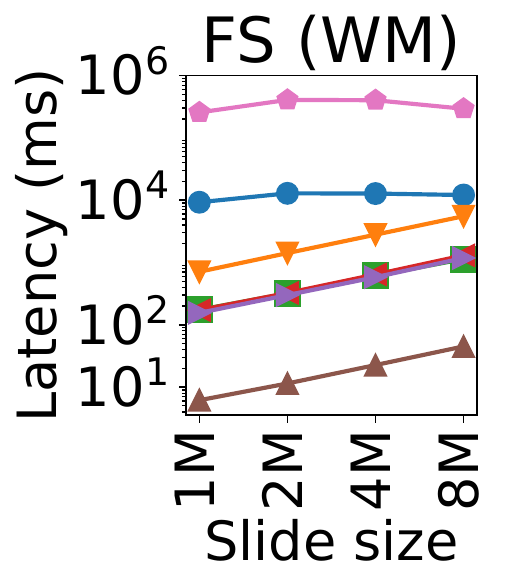}
    }
        \end{minipage}
        \begin{minipage}{0.32\textwidth}
            \resizebox{\textwidth}{!}{
    \includegraphics{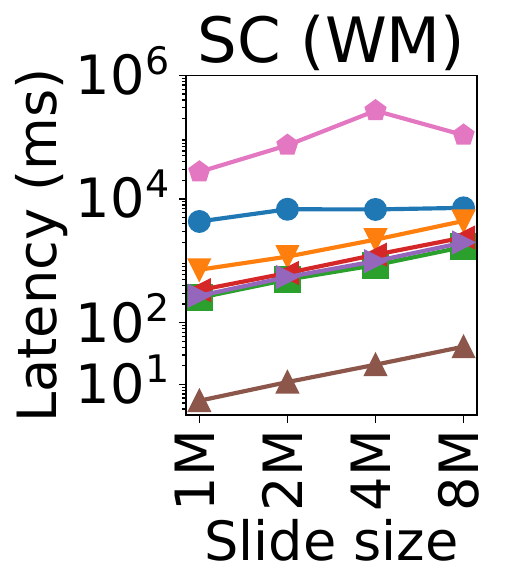}
    }
        \end{minipage}
    \end{minipage}
    \caption{Throughput, and P99 latency for query (Q) latency and  window management (WM) using slides of various sizes.}\label{fig:scale_slide_size}
\end{figure*}

\begin{figure*}
\centering
\resizebox{0.75\linewidth}{!}{
\includegraphics{figures/exp/scalability-throughput-legend.pdf}
}     
\begin{minipage}{0.29\linewidth}
    \resizebox{\textwidth}{!}{
        \includegraphics{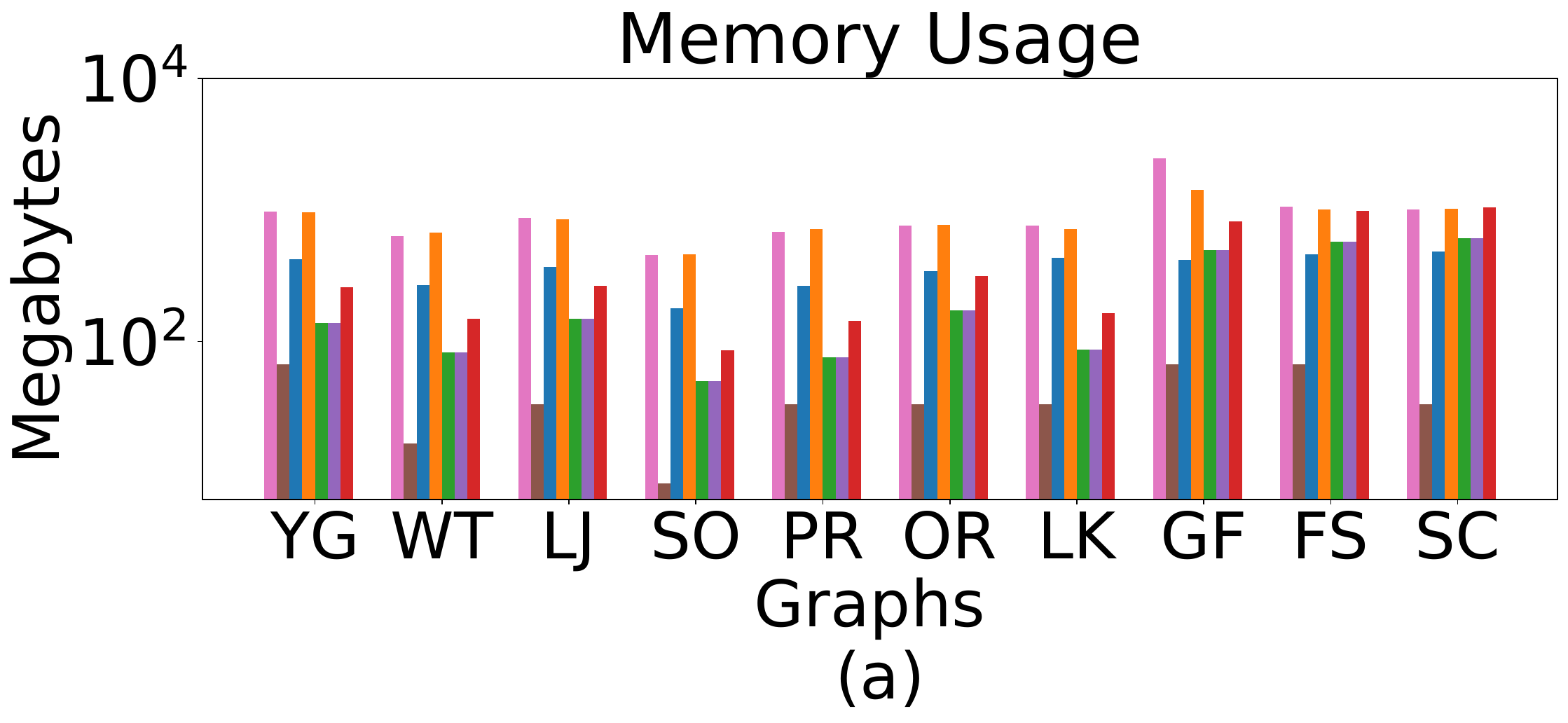}
    }
\end{minipage}
\begin{minipage}{0.69\linewidth}
    \begin{minipage}{0.49\linewidth}
    \begin{minipage}{0.32\textwidth}
        \resizebox{\textwidth}{!}{
\includegraphics{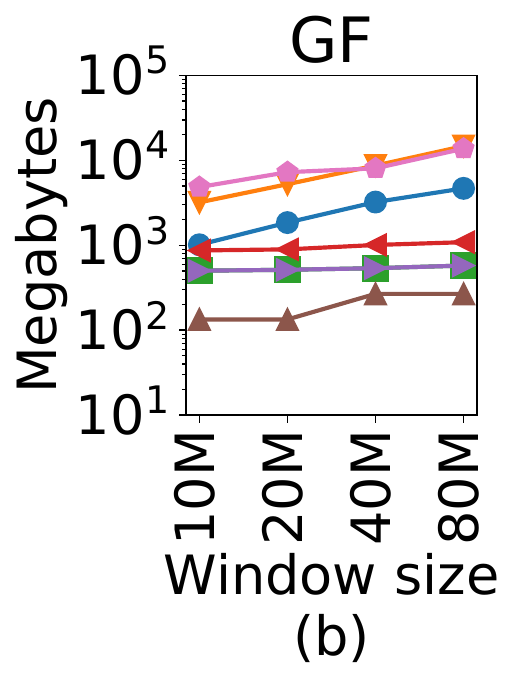}
}
    \end{minipage}
    \begin{minipage}{0.32\textwidth}
        \resizebox{\textwidth}{!}{
\includegraphics{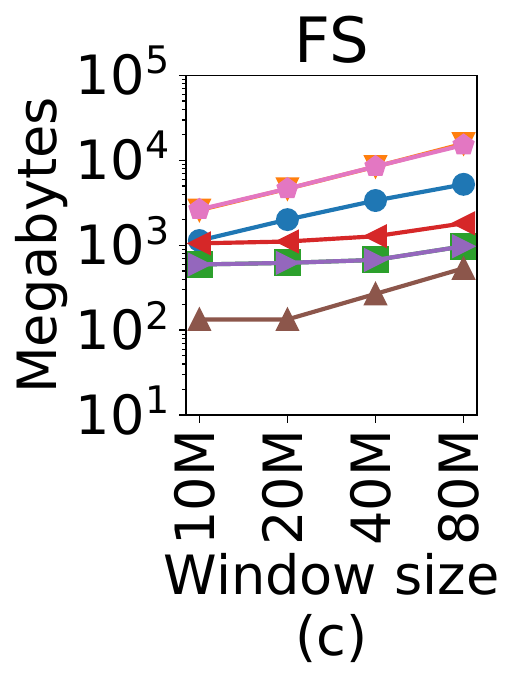}
}
    \end{minipage}
    \begin{minipage}{0.32\textwidth}
        \resizebox{\textwidth}{!}{
\includegraphics{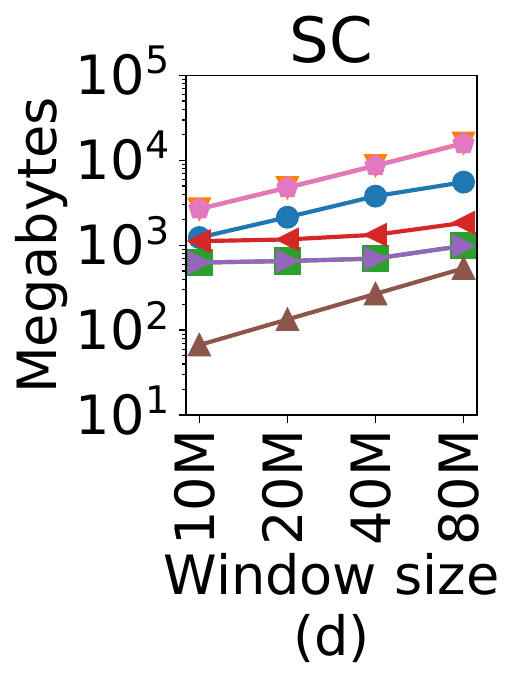}
}
    \end{minipage}
\end{minipage}
    \begin{minipage}{0.49\linewidth}
    \begin{minipage}{0.32\textwidth}
        \resizebox{\textwidth}{!}{
\includegraphics{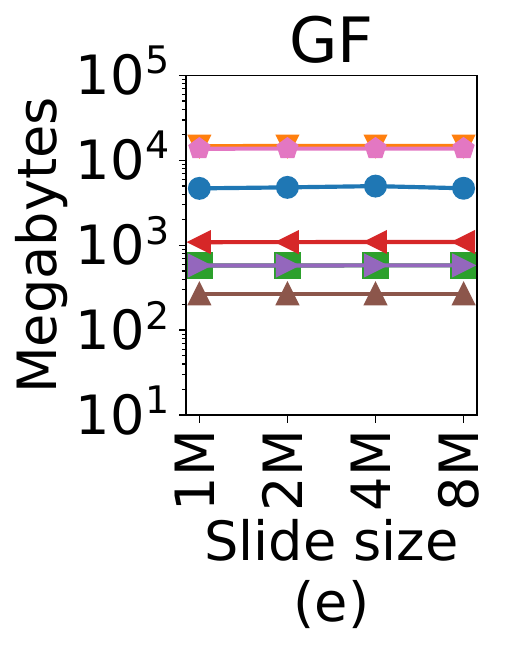}
}
    \end{minipage}
    \begin{minipage}{0.32\textwidth}
        \resizebox{\textwidth}{!}{
\includegraphics{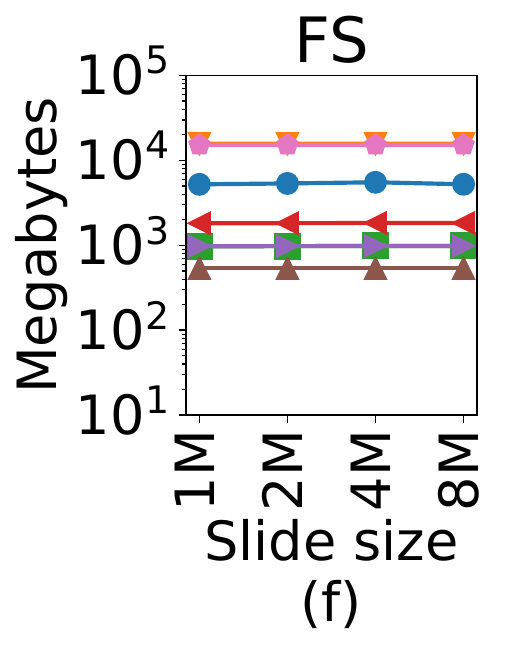}
}
    \end{minipage}
    \begin{minipage}{0.32\textwidth}
        \resizebox{\textwidth}{!}{
\includegraphics{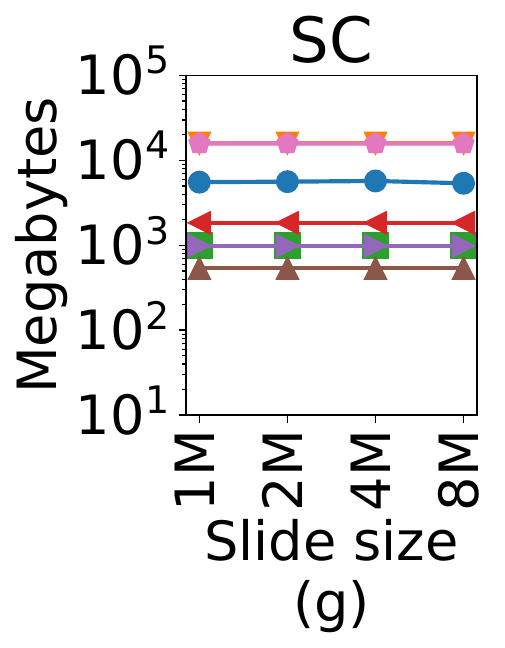}
}
    \end{minipage}
    \\
\end{minipage}
\end{minipage}
\caption{
Memory usage: (a) analyzes per graph; (b)-(d) assess window size impact; (e)-(g) evaluate slide size impact.
}\label{fig:per-memory}
\end{figure*}

\subsection{Impact of Window Sizes and Slide Sizes}\label{sec:scalability}
\textit{Window sizes}.
Figure \ref{fig:scale_window_size} presents our experimental results using a fixed slide interval of $1$M edges on average and a workload of $10^5$ queries, while varying window sizes. 
In general, the window size significantly impacts RWC and BIC, but has a lesser effect on the approaches based on spanning trees, including Vanilla D-Tree, MST D-Tree, OMST D-Tree, OMST S-Tree, and OMST LC-Tree. 
Overall, OMST D-Tree and OMST S-Tree demonstrate the best throughput and latency results across various window sizes.
RWC computes connected components from scratch for each window snapshot by scanning all the edges in each window snapshot. Consequently, larger window sizes result in higher computation latency, as evidenced by its increased query latency. This higher latency significantly decreases RWC’s throughput.
BIC recalculates its backward UFTs from scratch for every disjoint chunk of streaming edges containing $\alpha $ timestamps. As window size $\alpha$ increases, the corresponding latency also increases, particularly observable in its window management latency. Additionally, larger window sizes elevate the computation cost for query processing due to longer traversals across the larger backward and forward UFTs, as reflected in the increased query latency. This results in a throughput drop for BIC.
The impact of increasing window sizes on approaches utilizing spanning trees is minimal. While larger windows may result in spanning trees containing more vertices, these methods employ efficient techniques, such as linking the root of a smaller tree as a child of a larger tree when inserting a tree edge, which minimally increases tree depth. Consequently, the main operation—traversing the path from vertices to their roots—remains relatively unaffected by changes in window size, as shown by the query latency results.
Finally, MST D-Tree, OMST D-Tree, OMST S-Tree, and OMST LC-Tree consistently outperform Vanilla D-Tree in throughput and window management latency, underscoring the effectiveness of our framework in handling various window sizes.

\textit{Slide sizes}.
Figure \ref{fig:scale_slide_size} presents our experimental results using a fixed window size of $80$M edges on average and a workload of $10^5$ queries, while varying slide sizes. Increasing slide sizes leads to throughput improvements for both RWC and BIC, as larger slides result in less frequent query processing. Consequently, the operations that represent major performance bottlenecks—computing connected components from scratch for RWC and computing backward UFTs and traversing both backward and forward UFTs for BIC—are executed less often. 
Since slide sizes do not affect the number of edges within windows, their query latency generally remains unchanged.
For window management latency, BIC does not involve the deletion of expired edges, keeping its latency consistent. However, RWC requires removing expired edges from windows, which leads to an increase in window management latency.
For spanning tree-based approaches—Vanilla D-Tree, MST D-Tree, OMST D-Tree, OMST S-Tree, and OMST LC-Tree—the slide size does not impact the path length from vertices to their roots, so query latency remains stable. However, their window management latency increases due to the need to delete expired edges. While MST D-Tree, OMST D-Tree, OMST S-Tree, and OMST LC-Tree do not require replacement edge searches during edge deletions, they still involve constant effort to remove edges from both spanning trees and windows, resulting in increased computation times as slide sizes grow.
The window management latency differences between Vanilla D-Tree and MST D-Tree highlight the advantages of our MST framework, which eliminates the need for replacement edge searches. Furthermore, the performance distinction between MST-based and OMST-based approaches illustrates the benefits of the optimization techniques introduced in the OMST framework.
We note that increasing the slide size generally simplifies the problem of computing sliding window connectivity, as the number of overlapping window snapshots decreases. When the slide size equals the window size, the scenario transitions into a \textit{tumbling window}, which is typically easier to handle than sliding windows. As a result, larger slide sizes reduce the overall computational effort.

\subsection{Memory Usage}\label{sec:memory_udage}
Memory consumption results are presented in Figure \ref{fig:per-memory}: \ref{fig:per-memory}(a) details memory usage for each graph with a window size of 3M edges and a slide size of 150K edges; \ref{fig:per-memory}(b)-(d) illustrate memory usage with a fixed slide size of 1M edges across various window sizes; \ref{fig:per-memory}(e)-(f) depict memory usage for a fixed window size of 80M edges with varying slide sizes.
Overall, OMST D-Tree and OMST S-Tree use the least amount of memory among index-based approaches.
RWC consumes the least amount of memory because it recomputes connected components for each window, resulting in memory usage linear to the number of vertices in windows.
OMST D-Tree, OMST S-Tree, and OMST LC-Tree consume less memory than MST D-Tree, as they store only spanning trees, while MST D-Tree also stores non-tree edges, similar to Vanilla D-Tree.
OMST D-Tree and OMST S-Tree have identical memory usage, less than OMST LC-Tree, which requires extra pointers for its \texttt{access} operation. 
BIC, at least storing all streaming edges in each disjoint chunk that contains  all streaming edges in a window, generally uses more memory than OMST-based approaches. 
Memory usage increases with window size for all but the OMST framework approaches; slide size does not affect memory usage.
Although both RWC and the OMST framework approaches (OMST D-Tree, OMST S-Tree, and OMST LC-Tree) have the same space complexity of 
$O(n)$, where $n$ is the number of vertices in a window, their actual memory usage differs. 
In the OMST framework, only a parent pointer is maintained for each vertex in spanning trees. When deleting expired edges, if a vertex is no longer connected to any edges in the spanning tree, it is not removed. Deleting such vertices would require maintaining child pointers for each vertex, effectively doubling the storage requirements.
As a result, the OMST framework might store more vertices than RWC. 
However, in scenarios with large window sizes, the memory usage between the OMST framework and RWC becomes similar, as shown in Figures \ref{fig:per-memory}(b)-(d).
We note that maintaining child pointers for each vertex is straightforward and does not overly complicate the structure.

\section{CONCLUSION}
Index-based approaches for processing connectivity queries over sliding windows require not only efficient query processing but also timely index updates when new edges are inserted or expired edges are deleted. Existing methods fail to achieve both, leading to high latency and low throughput.
In this paper, we introduce a novel indexing framework that utilizes maximum spanning trees to overcome these limitations. Our framework leverages spanning trees for efficient query processing and updates them efficiently by eliminating the need to search for replacement edges when deleting expired edges, which is a major performance bottleneck in traditional spanning tree-based approaches.
We integrate various spanning tree techniques into our framework, from simple trees to more sophisticated structures like D-Trees and Link-Cut Trees. We also identify numerous optimization techniques to enhance their performance and reduce memory usage. 
Our comprehensive experimental results show that our methods reduce query latency by $1172\times$, window management latency by $13\times$, and improve throughput by $80\times$ compared to the most recent baseline. Against FDC approaches, we achieve a $458\times$ reduction in window management latency and $8\times$ better throughput, with similar query latency. Our framework also uses less memory and consistently excels across different settings, including query load, window size, and slide intervals.

\bibliographystyle{ACM-Reference-Format}
\bibliography{publications,bibliography}

\appendix
\section{Proof of Lemma \ref{lemma:replacement_edge}}
\begin{proof}
Let $\beta$ be the slide interval of a time-based sliding window. Consider window snapshot $\window_i$ with beginning timestamp $w_i.t_b$. 
When $\window_i$  transitions to $\window_{i+1}$, the expired edges (EEs) that need to be deleted from $\window_{i}$ are the following ones: $ EE = \{e \mid w_i.t_b \leq e.t < w_i.t_b + \beta \}$.
Consider an expired edge $e \in EE$ that needs to be deleted from $\window_i$. Assume there exists a replacement edge $e'\in \window_i$ for $e$. Since the spanning tree containing $e$ is an MST, we have $e'.t \leq e.t$; otherwise $e'$ would be a tree edge and $e$ a non-tree edge. 
This implies $e' \in EE$, meaning that $e'$ will also be deleted during deleting all edges in $EE$. 
\end{proof}

\section{Proof of Lemma \ref{lemma:min_edge}}
\begin{proof}
    The cycle property states that for any cycle $C$ in the graph, if the weight of an edge $e$ of $C$ is smaller than any of the individual weights of all other edges of C, then this edge cannot belong to an maximum spanning tree.
    Therefore, the minimum edge in the cycle $C$ containing $(u,v)$ cannot belong to any MST. 
\end{proof}

\section{Proof of Theorem \ref{theorem:mst_swc}}
\begin{proof}
    Lemma \ref{lemma:min_edge} guarantees the invariant of MSTs. Thus, searching for  replacement edges can be avoided according to Lemma \ref{lemma:replacement_edge}. Queries can be correctly processed based on the proper maintenance of spanning trees.
\end{proof}

\end{document}